%% file: LDF.tex
\documentclass[preprint,12pt,authoryear]{elsarticle}
\usepackage{amssymb,amsmath}        
\usepackage{amsthm}
\usepackage{mathtools}
\usepackage{bbm} 
\usepackage{xcolor}
\usepackage{caption}
\usepackage{subcaption}
\usepackage{pdflscape}
\usepackage{afterpage}
\usepackage{placeins}
\usepackage{algorithmic}
\usepackage{booktabs}
\usepackage{float}
\usepackage{url}
\usepackage{setspace}

\let\geq=\geqslant
\newtheorem{theorem}{Theorem}
\newtheorem{corollary}{Corollary}[theorem]
\numberwithin{equation}{section}  

\DeclareMathOperator{\E}{\mathbb{E}}

\begin{document}   

\def\spacingset#1{\renewcommand{\baselinestretch}%
{#1}\small\normalsize} \spacingset{1} 

\begin{frontmatter}

\title{A loss discounting framework for model averaging and selection in time series models}


\author[1]{Dawid Bernaciak\corref{cor1}
}
\ead{dawid.bernaciak@ucl.ac.uk}
\author[1]{Jim E. Griffin}

\cortext[cor1]{Corresponding author}
\address[1]{Statistical Science, University College London, Gower Street, London WC1E 6BT, U.K.}



\begin{abstract}
We introduce a Loss Discounting Framework for model and forecast combination which generalises and combines Bayesian model synthesis and generalized Bayes methodologies. We use a loss function to score the performance of different models and  introduce a multilevel discounting scheme which allows a flexible specification of the dynamics of the model weights.
This novel and simple model combination approach  can be easily applied to 
 large scale model averaging/selection, can handle unusual features such as sudden regime changes, and can be tailored to different forecasting problems. We compare our method  to both  established methodologies and state of the art methods for a number of macroeconomic forecasting examples. We find that the proposed method offers an attractive, computationally efficient alternative to the benchmark methodologies and often outperforms more complex techniques.
\end{abstract}


\end{frontmatter}


\input{parts/introduction.tex}

\input{parts/background.tex}

\input{parts/methodology.tex}

\input{parts/simulation.tex}

\input{parts/applications.tex}

\input{parts/discussion.tex}

\bibliographystyle{chicago}                
\bibliography{references}                     

\newpage
\appendix
\input{parts/appendices/technical_appendix.tex}

\input{parts/appendices/appendix_simulation.tex}

\input{parts/appendices/appendix_param_c.tex}


\input{parts/appendices/appendix_fx.tex}

\input{parts/appendices/appendix_usinflation.tex}
\end{document}

%% file: parts/introduction.tex
\section{Introduction}             
\label{section:intro}                  

Recent developments of econometric modelling and machine learning techniques combined with  increasingly easy access to vast computational resources and data has lead to a proliferation of forecasting models yielding either point forecasts or full forecast density functions. This trend has been met with a renewed  interest in tools that can effectively use  these different forecasts such as
model selection, or forecast combination, pooling or synthesis, {\it e.g.}, \citet{stock2004combination}, 
\citet{hendry2004pooling}, \citet{hall2007combining}, \citet{raftery2010online}, 
\citet{geweke2011optimal}
\citet{waggoner2012confronting}, \citet{koop2012forecasting}, \citet{billio2013time}, \citet{del2016dynamic}, \citet{yao2018using}, \citet{McAlinn_2019}, \citet{diebold2021aggregation}, \citet{li2023bayesian} to mention just a few. 
 \citet{Wang23} provide an excellent recent review of work in this area.




Combining forecasts from different models, rather than using a forecast from a single model is intuitively appealing and justified by improved empirical performance \citep[see \textit{e.g.}][]{granger69, stock2004combination}.
\citet{hendry2004pooling} suggested that  combining  point forecasts provides an insurance against poor performance by individual models which are  misspecified, poorly estimated or non-stationary.

In density forecasting, the superiority of a combination over single models is less clear. Bayesian model averaging (BMA) \citep{leamer1978specification} is simple and coherent approach to  weight forecasts in a combination
but may not be optimal under logarithmic scoring when the set of models to be combined is misspecified
\citep{diebold1991note}. 
 Since sets of models will usually not include the true data generating mechanism, this result has driven a substantial literature proposing alternatives to BMA.
 \citet{hall2007combining}  
 proposed
 a  logarithmic scoring rule for a time-invariant linear pool with weights on the simplex which leads to a 
forecast density combination that  minimises  Kullback-Leibler divergence to the true but unknown density. 
This idea has been developed to Bayesian estimation \citep{geweke2011optimal}, Markov switching weights
\citep{waggoner2012confronting}  and dynamic linear pools \citep{del2016dynamic, billio2013time}. These approaches often lead to better forecasting performance but at the cost of increased computational expense.
 A computationally cheaper alternative 
  directly adjusts the model weights from BMA to allow time-variation \citep{raftery2005using} leading to  Dynamic Model Averaging (DMA)  \citep{raftery2010online}, which uses an exponential discounting of Bayes factors with a discount/forgetting/decay\footnote{The terms discount/forgetting/decay factor are used interchangeably in this paper.} to achieve time-varying model weights. 
Performance can be sensitive to the discount factor and
\citet{koop2012forecasting} suggested using logarithmic score maximisation to find an optimal discount  factor for DMA. \citet{Koop2020}
applied this idea to model selection and developed Dynamic Model Learning (DML) method with an application to foreign exchange forecasting. 
Outside the formal Bayesian framework, \citet{diebold2021aggregation} suggested a simple average of the forecasts from a team of $N$ (or less) forecasters chosen using the average logarithmic scores
in the previous  $rw$-periods. 
This can be seen as a localised and simplified version of \citet{hall2007combining}.


Recently, \citet{McAlinn_2019} and \citet{mcalinn2020multivariate} proposed a broad theoretical framework called Bayesian Predictive Synthesis (BPS) which includes the  majority of proposed Bayesian techniques as special cases.
 They propose a novel forecast combination method using latent factor regression, cast as a Bayesian seemingly unrelated regression (SUR)
and showed better performance than  the BMA benchmark and an optimal linear pool. However, the approach can be computationally demanding with a large pool of models.
  \cite{tallman2022bayesian} use entropic tilting to expand the BPS framework to more  general aims than forecast accuracy (such as return maximisation in portfolio allocation).

This paper describes our Loss Discounting Framework (LDF) which extends DMA and DML to general loss function 
\citep[in a similar spirit to][]{tallman2022bayesian}, and more general discounting dynamics. 
A computationally efficient time-varying discounting scheme is constructed through a sequence of pools of meta-models, which starts with the initial pool. Meta-models at one layer are constructed by combining meta-models at the previous layer using a DMA/DML type rule with different discount factors. We show that LDF can outperform other benchmarks methods and is more robust to hyperparameter choice than DMA/DML in simulated data, and foreign exchange forecasting using  econometric fundamentals using a large pool of models.
 We also  show how tailoring the approach to constructing
 long-short foreign exchange portfolios can lead to  economic gains. A second example 
 illustrates the limitations of our methodology in US 
 inflation forecasting.

The paper is organised as follows. Section \ref{section:Background} presents some background leading into a description of the proposed methodology in Section \ref{section:extension}. In Section \ref{section:examples}, the performance of the LDF approach is examined in a simulated example 
and applications to foreign exchange and US inflation forecasting.
  We discuss our approach and set out directions for further research in Section \ref{section:discussion}. The code to reproduce our study is freely available from  
  \url{https://github.com/dbernaciak/ldf}.



%% file: parts/background.tex
\section{Background}
\label{section:Background}
It is common in Bayesian analysis \citep[][and references therein]{bernardo2009bayesian, yao2018using}  to distinguish three types of model  pools  $\mathcal{M} = \{M_1, M_2, \cdots, M_K\}$: 
 $\mathcal{M}$-closed -- the true data generating process is described by one of the models in $\mathcal{M}$ but is unknown to researchers; $\mathcal{M}$-complete -- the model for the true data generating process exists but is not in $\mathcal{M}$, which is viewed as a set of  useful approximating models; $\mathcal{M}$-open -- the model for the true data generating process is not in $\mathcal{M}$ and the true model cannot be constructed either in principle or due to a lack of resources, expertise etc.\footnote{\citet{clarke2013prediction} give, a slightly unusual, example of works of William Shakespeare as an $\mathcal{M}$-open problem. The works (data) has a true data generating process (William Shakespeare) but one can argue that it makes no sense to model the mechanism by which the data was generated.} Model selection based on BMA only converges to the true model in the $\mathcal{M}$-closed case 
 \citep[see {\it e.g.}][]{diebold1991note} and can perform poorly otherwise.

There are several reasons to believe that econometric problems are 
outside the $\mathcal{M}$-closed setting. 
Firstly, real-world forecasting applications often involve complex systems and the model pool will only include approximations at best. In fact, one might argue that econometric modellers have an inherent belief that the models they propose provide reasonable approximation to the data generating process even if certain process features escape the capabilities of the supplied methodologies.  Secondly, in many applications, the data generating process is not constant in time  \citep{del2016dynamic} and may involve regime changes and 
 considerable model uncertainty. For example, in the foreign exchange context, \citet{bacchetta2004scapegoat} proposed the scapegoat theory suggesting that investors display a rational confusion about the true source of exchange rate fluctuations. If an exchange rate movement is affected by a factor which is unobservable or unknown, investors may attribute this movement to some other observable macroeconomic fundamental variable. This induces regimes where different market observables might be more or less important.

These concerns motivate  a model averaging framework that is both, suitable for $\mathcal{M}$-complete (or even $\mathcal{M}$-open) situations and incorporates time-varying model weights.
We use $\pi_{t|s,k}$ to represent the weight of model $k$ at time $t$ using information to time $s$ and use the forecast combination density
\begin{equation}
p(y_{t}|y_s) = \sum_{k=1}^K \pi_{t|s, k} \, p_k(y_{t}|y_s)
\label{eqn::FDC}
\end{equation}
where $p_k(y_{t}|y_s)$ represents the forecast density of model $k$ at time $t$ using information $ y_1,\dots, y_s$, which we call the predictive likelihood.
DMA \citep{raftery2010online}, assumes that $s = t - 1$ and updates $\pi_{t+1|t, k}$
using the observation at time $t$, $y_t$, and a forgetting factor, denoted by $\alpha$, by the recursion
\begin{align}
\label{eq:rec1}
\pi_{t|t,k} &= \frac{\pi_{t|t-1,k} \, p_j(y_{t}|y_{t-1})}{\sum_{l=1}^K \pi_{t|t-1,l} \, p_l(y_{t}|y_{t-1})},\\
\label{eq:rec2}
\pi_{t+1|t,k} &= \frac{\pi_{t|t,k}^{\alpha} + c}{\sum_{l=1}^K \pi_{t|t,l}^{\alpha} + c},
\end{align}
where $c$ is a small positive number introduced to avoid model probability being brought to machine zero by aberrant observations\footnote{\citet{Yusupova2019} note that this constant is only present in the original work by \citet{raftery2010online} and then in the implementation by \citet{koop2012forecasting} but then subsequently dropped in further works, software packages and citations. They also notice that this constant has a non-trivial and often critical effect of the dynamics of weight changes. We comment on this aspect in \ref{appendix:B}.}.
The log-sum-exp trick is an alternative way of handling this numerical instability which would, at least in part, eliminate the need for the constant $c$. 
We leave the role of this parameter to further research.

The recursions in \eqref{eq:rec1} and \eqref{eq:rec2} amount to a closed form algorithm to update the probability that model $k$ is the best predictive model given information up to  time $t$, for forecasting at time $t$.
A model receives a higher weight if it performed well in the recent past and the discount factor $\alpha$ controls the importance that one attaches to the recent past. For example, if $\alpha=0.7$, the forecast performance 12 periods prior to the last one receives approximately $2\%$ of the importance of  the most recent observation. However, if $\alpha=0.9$, this importance is as high as $31\%$. Therefore, lower values of $\alpha$ lead to large changes in the model weights. In particular, $\alpha \rightarrow 0$ would lead to equal model weights and $\alpha = 1$ recovers the standard BMA.

DMA has been shown to perform well in econometric applications whilst avoiding 
the computational burden of calculating 
large scale Markov Chain Monte Carlo (MCMC) or sequential 
Monte Carlo  associated with methods such as \citet{waggoner2012confronting}. 
\citet{del2016dynamic} showed that DMA 
performed comparably 
to their novel dynamic prediction pooling method in forecasting inflation and output growth.
It was subsequently expanded and successfully used in econometric applications by \citet{koop2012forecasting}, \citet{koop2013large} and \citet{Koop2020}. In the first two papers the authors compare DMA for a few possible values of discount factors $\alpha$, whereas, in the latest paper the authors follow the recommendation of \citet{raftery2010online} to estimate the forgetting factor online in the context of Bayesian model selection.
We find that estimating the forgetting factor is key to performance of DMA as we will show in our simulation study and empirical examples. Our LDF provides a 
general approach by 
 combining multiple layers of discounting with time-varying discount factors to provide better performance and robustness to the hyperparameter choice.

%% file: parts/methodology.tex
\section{Methodology}
\label{section:extension}
Our proposed loss discounting framework (LDF) provides a method of updating time-varying model weights using flexible discounting of a general measure of model performance. The flexible discounting is achieved by defining
 layers of meta-models using the simple discount scheme in \eqref{eq:rec1} and \eqref{eq:rec2}. The approach can be used for both dynamic model averaging and dynamic model selection.
For example, we can define a pool of forecast combination densities by applying \eqref{eqn::FDC} with different values of the discount factor. We refer to the elements of this pool as meta-models. We can subsequently find the best meta-model average (or best meta-model) by again applying exponential discounting to past performance of the meta-models. This leads to an approach with two layers but clearly we could continue the process by defining a pool of meta-models at one layer by applying the forecast combination in \eqref{eqn::FDC} to a pool of meta-models at the previous layer. 

The method has two key features. The first key feature of the model averaging (selection) we develop is the ability to shrink the pool of the relevant models (show greater certainty across time in a single model) in times of low volatility and to encompass more models (display greater variation in model selection) when the volatility of the system is high. The second key feature is the use of a generalised measure of model performance which enables user to define the scores/losses which is directly connected with their final goal. As we show in the empirical study, aligning model scores to the final purpose leads to better performance.
\subsection{Loss Discounting Framework}

We first describe how a score can be used to generalize DMA and then describe our discounting scheme using meta-models in more detail.
 The score or loss (we will use these terms interchangeably) is defined for the prediction of an observation with predictive distribution $p$ and observed value $y$  and  denoted
$S(p, y)$. This measures the quality of the predictive distribution if the corresponding observed value is $y$. For a set of  $K$ models, we assume that the (one-step ahead) predictive distribution for model $k$ at time $t$ is $p_{k, t} = p_k(y_t | y_{t-1})$
 we define
 the log-discounted predictive likelihood 
for the $k$-th model at time $t$ using discount factor $\alpha$ to be
\[
\mbox{LDPL}_{t, k}(\alpha) = \sum_{i=1}^{t-1} \alpha^{i-1} S\left(p_{k, t_i}, y_{t-i}\right).
\]
We define a model averaged predictive density
\[
\sum_{k=1}^K w_{t|t-1, k}(\alpha)\, p_k(y_t | y_{t-1})
\]
where 
\[
\left(w_{t | t - 1, 1}(\alpha), \dots, w_{t | t - 1, K}(\alpha)\right) = \mbox{softmax}\left(\mbox{LDPL}_{t, 1}(\alpha), \dots, \mbox{LDPL}_{t, K}(\alpha)\right).
\]
 This generalizes the use of the logarithmic scoring in DMA.
 The use of scoring rules for 
Bayesian updating for parameters was pioneered by \cite{Bissiri16} (rather than inference about models in forecast combination) and is justified in a $\mathcal{M}$-open or misspecified setting.  \cite{loaiza2021focused}
extend this approach to econometric forecasting.
They both consider sums which are equally weighted ({\it i.e.} $\alpha = 1$ for layer 2). \cite{Miller19} provide a justification for using a powered version of the likelihood of misspecified models.

 Each meta-model is defined using a recipe for model or meta-model averaging/selection. We consider a specific type of such recipe which is based on exponential discounting of the scores with different discount factor from a set of possible values $\mathcal{S}_{\alpha} = 
\{\alpha_1, \dots, \alpha_M\}$. To lighten the notation, we write $w(m)$ and $\mbox{LDPL}(m)$ to denote weights and log-discounted predictive likelihoods evaluated at $\alpha_m$.

In the first layer, each model in the model pool is scored and the $i$-th meta-model 
is defined by applying either DMA or DML with discounting $\alpha_i$ and the weights defined above.

Then, to construct the second layer, the meta-models in the first layer are scored and the $i$-th meta-model is again defined by applying either DMA or DML with discounting $\alpha_i$ to these scores. This iterative process can be easily extended to an arbitrary number of layers.
We highlight two parallels between the methods used in LDF for time series models and concepts in Bayesian modelling. The first parallel is between the layers of meta-models in LDF and the use of hyperpriors in the Bayesian hierarchical models. Similarly to making a decision on the set up of hyperpriors in the hierarchical models LDF allows for varying depth and type of meta-model layers appropriate for the use case in question. 
We also draw analogy between the model selection versus the maximum a posteriori probability (MAP) estimate of the quantity, and model weights in model averaging versus full posterior distribution.



To provide a full description of the approach, 
 we  will write the forecast densities of the $K$ models 
 as $p_1^{(0)}(y_t | y_{t-1}),\dots, p^{(0)}_K(y_t | y_{t-1})$ to make notation consistent. At every other layer, we define predictive meta-models which are an average of (meta-)models at the previous layers. At the first layer, we directly use the forecast combination in \eqref{eqn::FDC} 
\[
p^{(1)}_m(y_t | y_{t-1}) = \sum_{k=1}^{K} w^{(1)}_{t | t - 1, k}(m) \,p^{(0)}_{k}(y_t | y_{t-1}) 
\]
and, for $n\geq 2$, we apply 
 \eqref{eqn::FDC} to the $M$ meta-models specified 
 at the previous layer,
\[
p^{(n)}_m(y_t | y_{t-1}) = \sum_{k=1}^{M} w^{(n)}_{t | t - 1, k}(m) \,p^{(n-1)}_{k}(y_t | y_{t-1}).
\]
To define the weights $w^{(n)}_{t | t - 1, k}$, we extend
 the log-discounted predictive likelihood 
for the $k$-th (meta-)model at the $n$-th layer at time $t$ using discount factor $\alpha_m$ to be
\begin{equation}
\mbox{LDPL}^{(n)}_{t, k}(m) = \sum_{i=1}^{t-1} \alpha_m^{i-1} S\left(p^{(n)}_k, y_{t-i}\right).
\end{equation}

The weights in layer $n$ are constructed using either softmax\footnote{$\mbox{softmax}(a_1, \dots, a_J) = \left(\frac{\exp\{a_1\}}{\sum_{j=1}^J \exp\{a_j\}}, \dots, 
 \frac{\exp\{a_J\}}{\sum_{j=1}^J \exp\{a_j\}}\right)$}  (to give a form of (meta-)model averaging) or argmax (to give a form of (meta-)model selection). We use the notation $L_n$ to be represent this operation in the $n$-th layer which can either take the value $s$ (softmax) or $a$ (argmax). If $L_n = s$, 
\[
\left(w^{(n)}_{t | t - 1, 1}(m), \dots, w^{(n)}_{t | t - 1, K}(m)\right) = \mbox{softmax}\left(\mbox{LDPL}^{(n-1)}_{t, 1}(m), \dots, \mbox{LDPL}^{(n-1)}_{t, K}(m)\right)
\]
if $n = 1$, or
\[
\left(w^{(n)}_{t | t - 1, 1}(m), \dots, w^{(n)}_{t | t - 1, M}(m)\right) = \mbox{softmax}\left(\mbox{LDPL}^{(n-1)}_{t, 1}(m), \dots, \mbox{LDPL}^{(n-1)}_{t, M}(m)\right)
\]
if $n \geq 2$.

If $L_n = a$,
\[
w^{(n)}_{t | t - 1, k}(m) =
\left\{\begin{array}{cc}
 1 & k = k^{\star}(m)\\
 0 & k\neq k^{\star}(m)
 \end{array}\right.
\]
where 
\[
k^{\star}(m) =  \mbox{argmax}\left(\mbox{LDPL}^{(r-1)}_{t, 1}(m), \dots, \mbox{LDPL}^{(r-1)}_{t, K}(m)\right)
\]
if $n=1$ or, if $n\geq 2$,
\[
k^{\star}(m) =  \mbox{argmax}\left(\mbox{LDPL}^{(r-1)}_{t, 1}(m), \dots, \mbox{LDPL}^{(r-1)}_{t, M}(m)\right).
\]
The $N$-layer LDF with score $S$ and
with choice $L_n$ (equal to $s$ or $a$) at layer $n$  will be written 
$\mbox{LDF}^N_{L_1L_2\dots L_N}(S)$.

The scheme only needs  a single discount factor to be chosen in the final meta-model layer. This parameter might be set by an expert or calculated on a calibration sample if the data sample is sufficiently large to permit a robust estimation. In LDF, We refer to the discount factor in the final meta-model layer as $\alpha$.

As well as defining a model combination at each layer, $\mbox{LDF}^N_{L_1L_2\dots L_N}(S)$ also leads to a discount model averaging of the initial model set for any $N$ since
\begin{align}
p^{(N)}_m(y_t | y_{t-1}) &= \sum_{k_N=1}^M w^{(N)}_{t | t - 1, k_N}(m) \,p^{(N - 1)}_{k_N}(y_t | y_{t-1}) \\
 &=  \sum_{k_1=1}^K \left[\sum_{k_2=1}^M \dots \sum_{k_N=1}^M
w^{(N)}_{t | t - 1, k_N}(m) \prod_{p=1}^{N-1} w^{(p)}_{t | t - 1, k_p}(k_{p+1}) \right]
 \,p_{k_1}^{(0)}(y_t | y_{t-1}).
 \label{ldf-recursive}
\end{align}

Given this set up the models and meta-models are either averaged by using the softmax function
 or selected by using the argmax function applied to the log-discounted predictive likelihood. 

\subsection{Special cases}
\subsubsection{Dynamic Model Averaging}
The updates of the Dynamic Model Averaging weights in \eqref{eq:rec2} correspond to passing $\mbox{LDPL}_{t,1}^{(0)}, \dots, 
 \mbox{LDPL}_{t,K}^{(0)}$ with the logarithmic scoring function through the softmax function. In DMA we only have one level of discounting where $p_k(y_t | y_{t-1})$ are the different forecaster densities. Therefore, we could denote DMA as $\mbox{LDF}_{\mbox{s}}^1$ where the superscript indicates a single level of loss discounting and the $\mbox{s}$ subscript indicates the use of the softmax function.

\subsubsection{Dynamic Model Learning}
\label{ss:DML}
Dynamic Model Learning (DML) \citep{Koop2020} provides a way to optimally choose a single  discount factor for the purposes of model selection. In DML the 
 logarithmic scoring function $S\left(p^{(0)}_k, y_{t-i}\right)$ are passed through an argmax function to select the best model. We could refer to DML as $\mbox{LDF}_{\mbox{a},\mbox{a}}^2$ with with the second layer of meta-models prior restricted to a single point on the grid, namely $\mathcal{S}_\alpha=\{ 1\}$ for $n=2$. 

A similar idea for model averaging using the softmax function for selection an ensemble of parameters $\alpha$, was developed in \citet{zhao2016dynamic}.

\subsubsection{Two-Layer Model Averaging/Selection within Loss Discounting Framework}
\label{subsection:2lvldf}
The Loss Discounting Framework allows us to describe more general set-ups for discounting in forecast combination, such as these models with two or more meta-model levels. In this paper we focus on LDF with two layers of meta-models such as $\mbox{LDF}_{\mbox{s}, \mbox{a}}^2$, $\mbox{LDF}_{\mbox{a}, \mbox{s}}^2$, $\mbox{LDF}_{\mbox{a}, \mbox{a}}^2$ and $\mbox{LDF}_{\mbox{s}, \mbox{s}}^2$, as well as, the limiting cases such as $\mbox{LDF}_{\mbox{s}\cdots \mbox{s}}^{\infty}$.
 In contrast to DMA and DML having two (with $\alpha \neq 1$) or more layers of meta-models makes the discount factors in the other layers time dependent which, as we show in the next sections, leads to an improved performance of model averaging and selection.

In terms of computation time our proposed algorithm is very fast as it just relies on simple addition and multiplication. This is an advantage over more sophisticated forecasts combination methods when the time series is long and/or we would like to incorporate a large (usually greater than 10) number of forecasters. 

As mentioned before, $\mbox{LDF}_{\mbox{a},\mbox{a}}^2$ is a generalised version of DML presented in \citet{Koop2020} where implicitly the authors suggest $\alpha=1$, i.e, all past performances of the forgetting factors are equally weighted. In the limit $\alpha \rightarrow 0$ we would choose the discount factor $\alpha$ which performed best in the latest run, disregarding any other history. The $\mbox{LDF}_{\mbox{a},\mbox{s}}^2$ specification is a hybrid between model selection and model averaging. The first layer performs the model selection for each discount factor, the second layer performs the model averaging for the discount factors. Therefore, for each discount factor we select a single model but then we take a mixture of discount factors which results in a mixture of models.

\subsection{Properties of LDF as $N\rightarrow\infty$}

It is natural to consider the impact of additional layers in an LDF model. 
If we use either the softmax or the argmax at all layers, the weights for each model converge as $N\rightarrow\infty$ and so adding more layers has a diminishing effect on the sequence of predictive distributions.
Intuitively, for the softmax functions, we have a diminishing impact on the final result as we take weighted averages of the weighted averages of the models, and for the argmax/model selection the LDF approach settles on a single model for any discount factor in the final layer. 
The detailed and rigorous proofs are provided in the technical \ref{appendix:X}. 
 We demonstrate in the empirical sections that the sequence converges to a predictive distribution which is often the best or nearly best performing set up of the LDF framework.

\subsection{Comments}


Low variation in LDPL across time leads to model weight concentration on fewer models and higher variation in LDPL leads to the opposite, the model weights are more evenly spread. This is because in a presence of high variation in LDPL the lower discount factors will be preferred and hence the faster forgetting which accommodates the regime switching. 

If one believes that the data generating process (DGP) is present in the pool, LDF will not perform as well as BMA which will asymptotically converge to the right model quicker than LDF. Conversely, if the DGP is not among the models in the pool, LDF adapts by adjusting the weights of the models over time to approximate the DGP.



Following the argument in \cite{del2016dynamic} to interpret DMA in terms of a Markov switching model, our extension allows a time-varying transition matrix, {\it i.e.} $Q_t=(q(t)_{kl})$. The gradual forgetting of the performance of the discount factor $\alpha$ allows for a change of optimal discount factor when the underlying changes in transition matrix are required. However, we also show that our two-layer model specification outperforms the standard DMA model even when the transition matrix is non-time-varying. This point will be further illustrated in \ref{appendix:A}.

%% file: parts/simulation.tex
\section{Examples}
\label{section:examples}

Our methodology is best suited to data with multiple regime switches with a potentially time-varying transition matrix. As such, it is particularly useful for modelling data such as inflation levels, interest or foreign exchange rates. We illustrate our model on a simulated example and two real data examples. The supplementary materials for our examples are given in \ref{appendix:A}, \ref{appendix:B}, 
\ref{appendix:D} and \ref{appendix:E}.

 We compare examples of our LDF to several popular model averaging methodologies. The approaches used are
\begin{itemize}
\item Multi-layer LDF - 2 hyperparameters, i.e., $\alpha$, $c$;
\item BMA - 0 hyperparameters;
\item DMA - 2 hyperparameters, i.e., $\alpha$, $c$
\item BPS \citep{McAlinn_2019} - 5 hyperparameters, i.e., $\beta$ discount factor for state evolution matrices, $\delta$ discount factor for residual volatility,  $n_0$ prior number of degrees of freedom, $s_0$ prior on BPS observation variance, $R_0$ prior covariance matrix of BPS coefficients;
\item best N-average \citep{diebold2021aggregation} - 2 hyperparameters, i.e., $N$ number of models, rolling window length $rw$.
\item DeCo \citep{billio2013time} - 5 hyperparameters (but defaults and online estimation options are available).
\end{itemize}

We evaluate the performance of the models by calculating the out-of-sample mean log predictive score (MLS)
\[
\mbox{MLS} = \frac{1}{T-s}\sum_{t=s+1}^T \log p(y_t\vert y_1,\dots, y_{t-1}),
\]
 and log predictive density ratios (LPDR) for a chosen reference model $m^*$
\[
\mbox{LPDR}(\tau) = \sum_{t=s+1}^\tau \log \{ p_{m^*}(y_t\vert y_1,\dots, y_{t-1}) / p_{LDF}(y_t\vert y_1,\dots, y_{t-1})\},
\]
where $y_1,\dots,y_s$ are the observations for a calibration period and $T$ is the total number of observations and $p_{LDF}$ correspond to the selected LDF model.

\subsection{Simulation study}             
\label{section:simulation} 

The data generating process (DGP) of \citet{diebold2021aggregation} is
\begin{align}
y_t &= \mu_t + x_t + \sigma_y \epsilon_t, \quad \epsilon_t \sim N(0,1),\\
x_t & = \phi_x x_{t-1} + \sigma_x v_t, \quad v_t \sim N(0,1),
\end{align}
where $y_t$ is the variable to be forecast, $x_t$ is the long-run component of $y_t$, $\mu_t$ is the time-varying level (in \citet{diebold2021aggregation} set to $0$). We can interpret $\mu_t$  as a piecewise-constant deterministic signal with a finite state space that accounts for regime switches. The error terms are all i.i.d and uncorrelated. It is assumed that the data generating process is known to each forecaster apart from the level component $\mu_t$.
Each individual forecaster $k$ models $x_t$ with noise and applies different level $\eta_{k}$ to $y_t$:
\begin{align}
z_{kt} &= x_t + \sigma_{tk} \nu_{kt}, \quad \nu_{kt} \sim N(0,1),\\
\tilde{y}_{kt} &= \eta_{k} + z_{kt} + \sigma_y \epsilon_t, \quad \epsilon_t \sim N(0,1).
\end{align}
Notice that the individual forecasters' levels are not time varying. This emulates a situation where forecasters have access to different sets of information and/or models which might guide a different choice of level. It emulates an $\mathcal{M}$-complete or even $\mathcal{M}$-open setting where no forecaster is right at all times.

In contrast to \citet{diebold2021aggregation}, we allow the variable $y_t$ to have multiple regime switches.  The settings are as follows: $\phi_x=0.9$, $\sigma_x = 0.3$, $\sigma_y = 0.3$, $\sigma_{tk} = 0.1 \, \forall k$, $K=20$, $T=2001$, $\eta_{k} = -2 + 0.2105(k-1), \, k=1,\ldots, K$ and finally:
\begin{align*}
\mu_t=
\begin{cases}
    0,& \text{for } t\in [0,49] \bigcup [200,399] \bigcup [800,849] \bigcup [970,979] \\& \bigcup [1000,1049] \bigcup [1600,1650] \bigcup [1700,2001]\\
    1, & \text{for } t\in [100,150] \bigcup [900,949] \bigcup [960,969] \bigcup [990,999] \\& \bigcup [1050,1099] \bigcup [1200,1599] \bigcup [1700,1749]\\
    -1, & \text{otherwise.}
\end{cases}
\end{align*}
More examples are discussed in \ref{appendix:A}, where we draw the levels from a Markov switching models 10 times.
For LDF we set $S_\alpha = \{1.0,\linebreak[1] 0.99,\linebreak[1] 0.95,\linebreak[1] 0.9,\linebreak[1] 0.8,\linebreak[1] 0.7,\linebreak[1] 0.6,\linebreak[1] 0.5,\linebreak[1] 0.4,\linebreak[1] 0.3,\linebreak[1] 0.2,\linebreak[1] 0.001\}$ and $c=10^{-20}$ similarly to \citet{koop2012forecasting}. 

\begin{figure}[!tb]    
\centering           
\includegraphics[width=0.8\textwidth, angle=0]{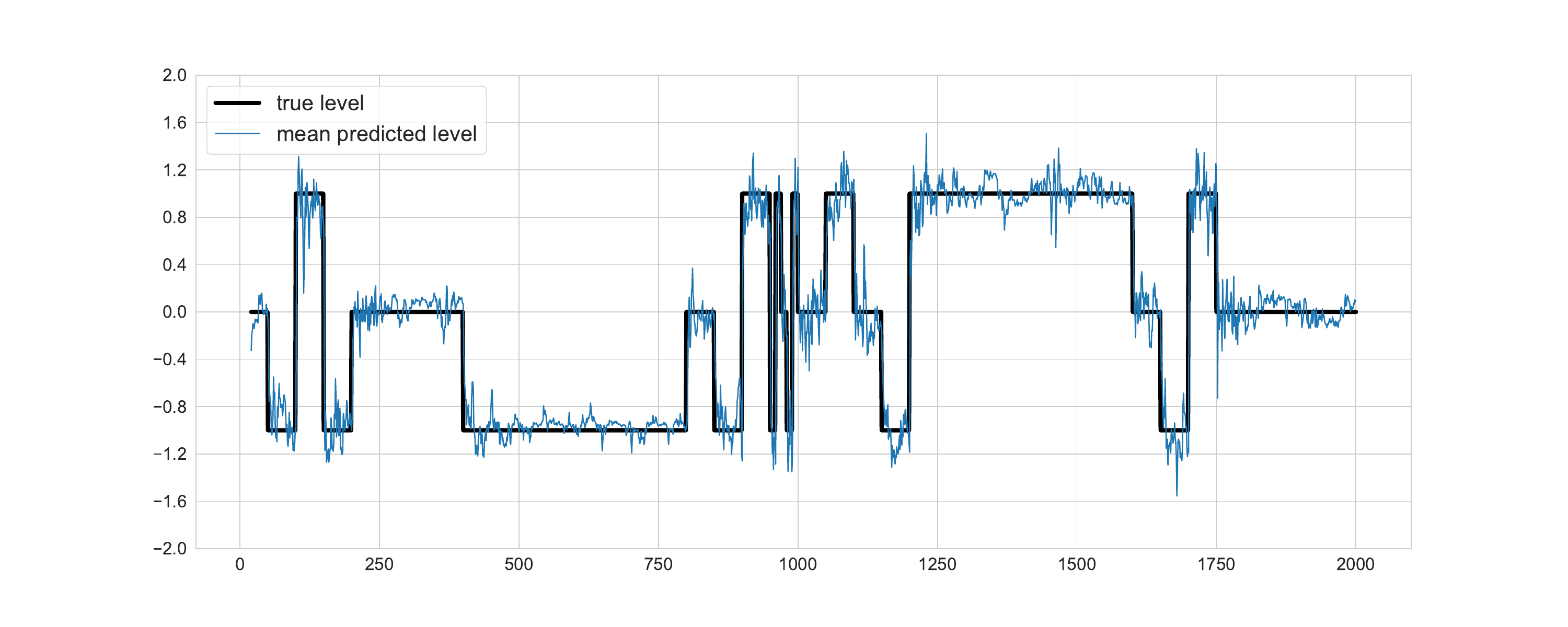}  
\vspace*{-0.25cm}    
\caption{Simulation -- True data generating process mean and mean predicted level according to $\mbox{LDF}_{\mbox{s},\mbox{s}}^2$.}  
\label{fig:example1}               
\end{figure}

In Figure \ref{fig:example1} we present how synthesised agent forecast level of $\mbox{LDF}_{\mbox{s},\mbox{s}}^2$ adjusts to the mean levels implied by the DGP. We can see that the model is very reactive to the mean predicted level following the true DGP mean closely with only a small time lag.

All results\footnote{Except BPS for which we performed only 1 run due to computational cost.} are based on 10 runs, where the levels were fixed but the random numbers regenerated. The standard Bayesian model averaging (MLS = -4.34) fared poorly since it quickly converged to the wrong model. DeCo (MLS=-0.57) was adapted to output 39 quantiles from which we calculated the log scores\footnote{39 quantiles in increments of 0.025. We used the default setting in DeCo package with $\Sigma=0.09$ (matching our DGP), and with learning and parameter estimation. The quantiles indicated that the output can be well approximated by the normal distribution.} did not manage to cope well with abrupt level changes in our numerical example, overestimated the variance which lead to poorer scores. BPS (MLS = -0.73)  with normal agent predictive densities\footnote{We used the original set of parameters (adjusted $\beta=0.95$ and $\delta=0.95$ to get better results as proposed by the authors of the paper but adjusted the prior variance to match the $\sigma_y$ parameter. The model was run for 5000 MCMC paths with 3000 paths burnin period. All other runs of BPS (which achieved worse results) are detailed in \ref{appendix:A})
} performed better than BMA but struggled to quickly adjust to the regime changes which resulted in low log-scores at the change points. The N-average method performed better (we chose rolling-window of 5 observations which performed best), with an MLS of -0.52 for $N=3$ and $N=4$, than BMA and BPS and similarly to the standard DMA method of \citet{raftery2010online}.

\begin{figure}[!tb]
     \centering
     \begin{subfigure}[b]{0.29\textwidth}
         \centering
         \includegraphics[width=\textwidth]{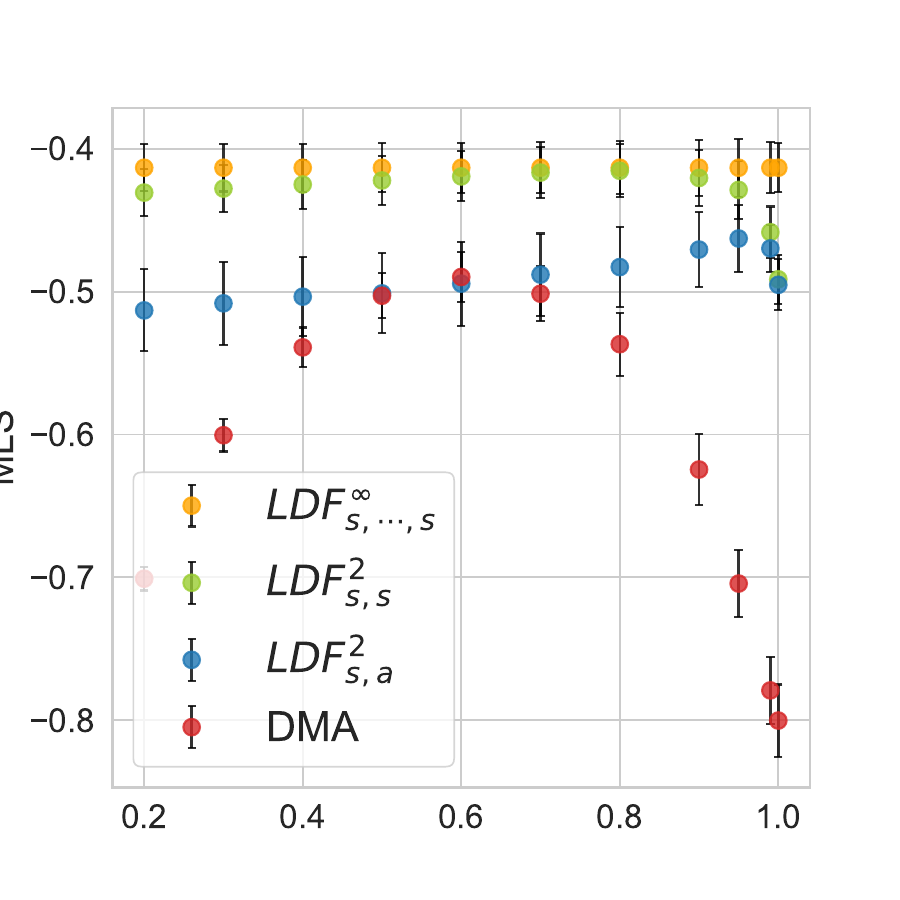}
         \caption{MLS versus values of $\alpha$.}
         \label{fig:simulation_hyper}
     \end{subfigure}
     \hfill
     \begin{subfigure}[b]{0.7\textwidth}
         \centering
         \includegraphics[width=\textwidth, trim={3cm 1cm 0 0},clip]{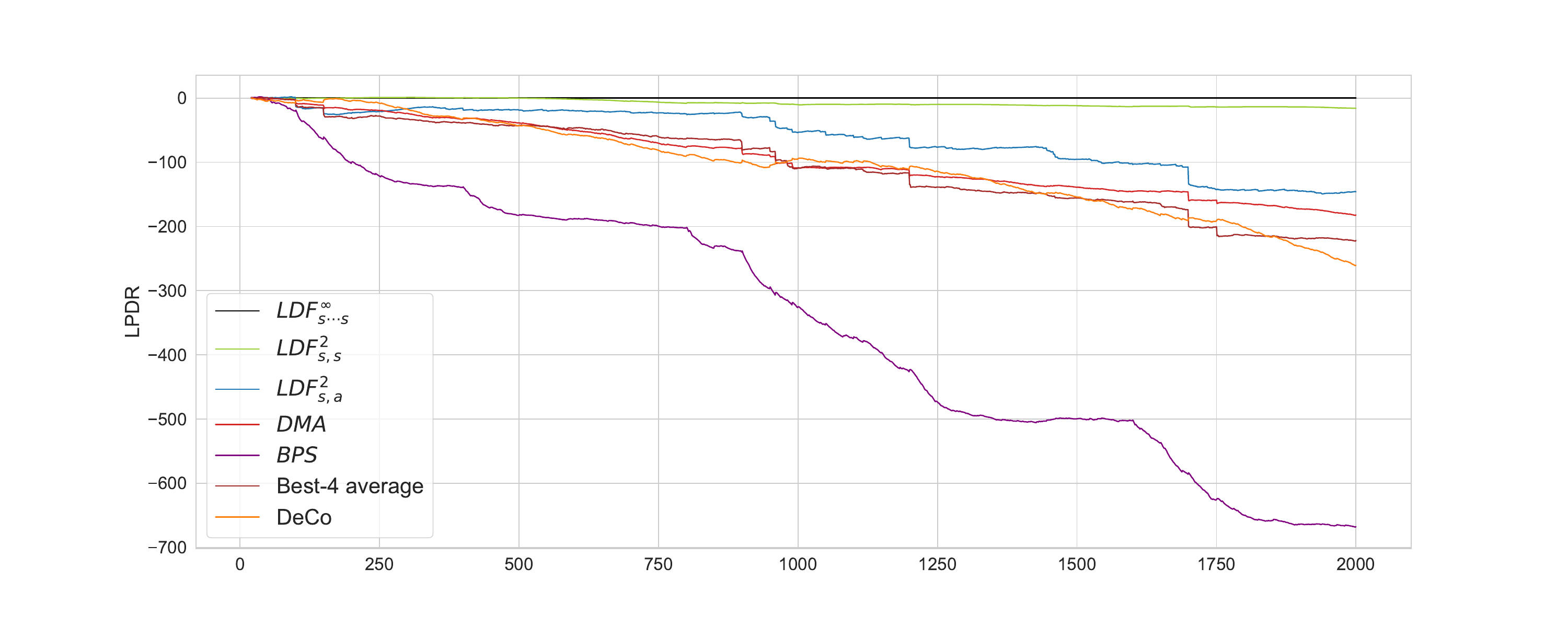}
         \caption{LPDR of the competing models.}
         \label{fig:simulation_lpdr}
     \end{subfigure}
     \caption{Simulation -- a) The MLS versus values of $\alpha$ for LDF and $\alpha$ for DMA in the x-axis. The error bars correspond to the standard deviation of MLS over 10 runs. b) LPDR of the competing models with calibration period of 250.}
\end{figure}

Crucially, we note that DMA's performance varies significantly depending on the hyperparameter choice, whereas the multilayered LDF methods' performance does not. This is clearly illustrated in Figure \ref{fig:simulation_hyper}. One could adopt various strategies in trying to find the hyperparameters. The most basic one would be based on tuning the hyperparameter on the calibration period and keeping the parameter constant thereafter. In this case, for example, if we set the calibration period to 250 the methods choose discounts as: DMA 0.5 (MLS = -0.50); $\mbox{LDF}_{\mbox{s},\mbox{s}}^2$ 0.6 (MLS = -0.42); $\mbox{LDF}_{\mbox{s},\mbox{a}}^2$ 0.7 (MLS = -0.49). For comparison the stable state $\mbox{LDF}_{\mbox{s},\cdots,\mbox{s}}^{\infty}$ achieves MLS = -0.41. The non-LDF model averaging models, namely BPS, DeCO and best N-average, were tuned to achieve best performance to the entire sample a posteriori in contrast to LDF models where we select a single configuration based on the initial sample of 250 observations. Another strategy could be based on selecting the best discount factor at each time step (online) based on the expanding window, potentially exponentially weighted - this boils down to an LDF approach with an additional argmax layer. In this case DMA simply becomes $\mbox{LDF}_{\mbox{s},\mbox{a}}^2$ and as shown can lead to better results. Even better results and more robustness can be achieved using $\mbox{LDF}_{\mbox{s},\mbox{s}}^2$ where a mix of discount factors is being used.

In Figure \ref{fig:simulation_lpdr} we present the LPDR for the tested models against $\mbox{LDF}_{\mbox{s},\cdots,\mbox{s}}^{\infty}$. The LDF models (including DMA) generally performed better, however, the results suggest that the 2-layer LDF, which can weight multiple discount factors model, is more robust to abrupt changes in the level than the other methods.
 
\begin{figure}[!tb]    
\centering           
\includegraphics[width=0.8\textwidth, angle=0]{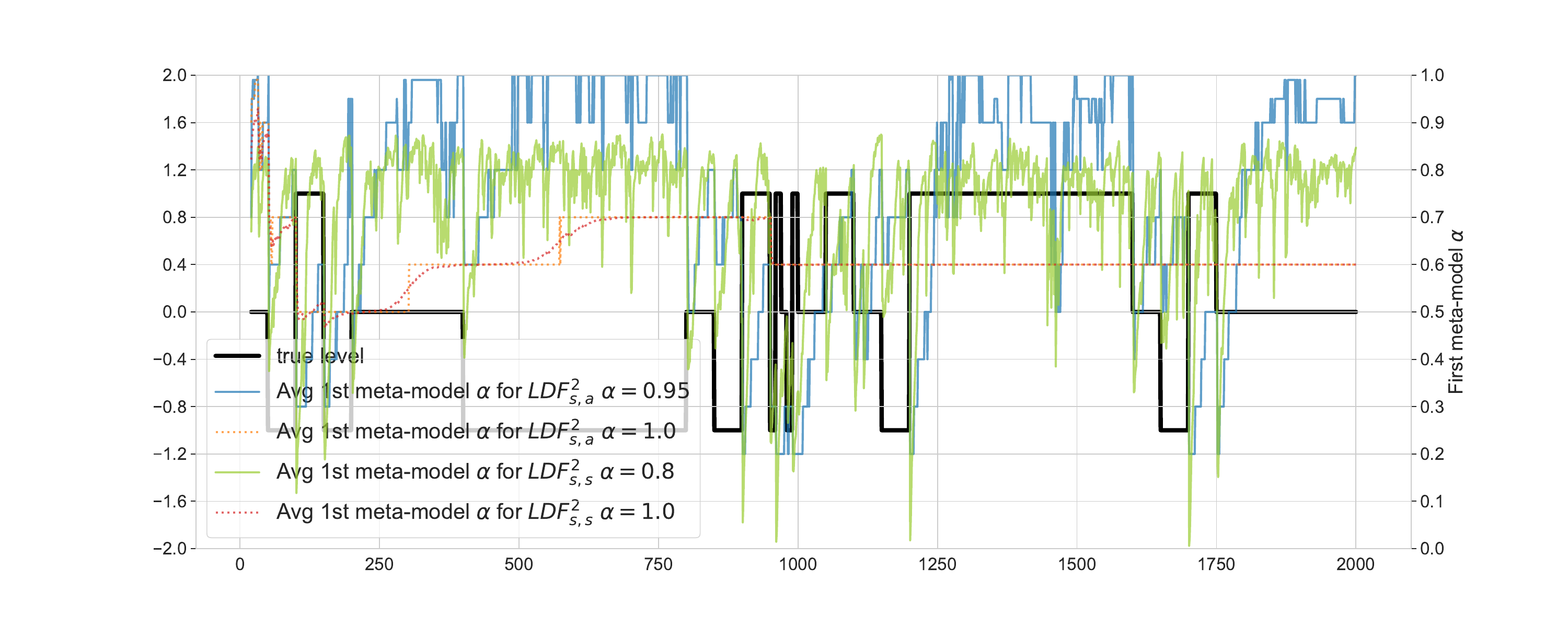}  
\vspace*{-0.25cm}    
\caption{Simulation -- Comparison of the average $\alpha$ parameters in the first meta-model layer for $\mbox{LDF}_{\mbox{s},\mbox{a}}^2$ model with $\alpha=0.95$ versus $\alpha=1$ as well as $\mbox{LDF}_{\mbox{s},\mbox{s}}^2$ with $\alpha=0.8$. We observe more dynamic adaptation of discount parameter in the fist meta-model layer when the final $\alpha<1$.}         
\label{fig:alpha}               
\end{figure}  

Figure \ref{fig:alpha} show how the average parameter $\alpha$ in the first meta-model layer dynamically changes using $\mbox{LDF}_{\mbox{s},\mbox{a}}^2$ with $\alpha=0.95$ and $\mbox{LDF}_{\mbox{s},\mbox{s}}^2$ with $\alpha=0.8$.
 It is close to $1$ in periods of stability and closer to 0 in times of abrupt changes. In comparison, for $\alpha=1$ the average parameter $\alpha$ is the  first meta-model layer is rather stable, oscillating around 0.6. As mentioned before, this variation in parameter $\alpha$ might be beneficial since the lower the $\alpha$ parameter more models will be taken into consideration and the final outcome might show more uncertainty. Additionally, a lower parameter $\alpha$ facilitates the ability to quickly re-weight the models to adapt to the new regime. Whereas, in the times of stability it might be better to narrow down the meaningful forecasts to a smaller group by increasing the parameter $\alpha$. This illustrates how the two layer model provides useful flexibility in the discount factors in the first meta-model layer. Another observation from Fig.\ \ref{fig:alpha} is concerning the average values of discount parameters $\alpha$ in the first layer across time. For $\mbox{LDF}_{\mbox{s},\mbox{a}}^2$ the average $\alpha$ in the first payer for $\alpha=0.95$ in the second layer is 0.75 and for $\mbox{LDF}_{\mbox{s},\mbox{s}}^2$ with $\alpha=0.8$ in the last layer it is 0.71. Whereas the average $\alpha$ for both LDF models with $\alpha=0.1$ in the last layer is 0.61.

%% file: parts/applications.tex
\input{parts/subparts/fx_2}


\input{parts/subparts/usinf.tex}

%% file: parts/subparts/fx_2.tex

\subsection{Foreign Exchange Forecasts}             
\label{ss:fx}    
We consider exchange rate forecasting 
\citep[see][for a comprehensive review]{ROSSI2013}. The random walk is a typical benchmark 
as it corresponds to the claim that the exchange rates are unpredictable but \cite{ROSSI2013} argues that 
economic variables can have 
time-varying predictive power.
\citet{Koop2020} consider
 exploiting this predictive ability
using DML with a pool of Time-Varying Parameter Bayesian Vector Autoregressive (TVP-BVAR) models with different subsets of economic fundamentals.
We closely follow their set-up. \ref{appendix:tvp-var} and \ref{appendix:fx-model-param}  describe the model.\footnote{
Following \citet{koop2013large} we adopt an Exponentially Weighted Moving Average (EWMA) estimator for the measurement covariance matrix to avoid the need for the posterior simulation for multivariate stochastic volatility. This is different than \citet{Koop2020} who use the approximation derived by \citet{triantafyllopoulos2011time}}

We use a set of G10 currencies: Australian dollar
(AUD), Canadian dollar (CAD), euro (EUR), Japanese yen (JPY), New
Zealand dollar (NZD), Norwegian krone (NOK), Swedish krona (SEK), Swiss
franc (CHF), pound (GBP) and US dollar (USD).
All currencies are expressed in terms of the amount of dollars per unit of a foreign currency, i.e. the domestic price of a foreign currency.
The data is monthly\footnote{If month end data was not available, it was substituted with the beginning of the month data or monthly average.
These substitutions were unavoidable for some of the data in the 1980s. See Appendix \ref{a:data} for more details.}
and runs from November 1989 to July 2020. This is a more up-to-date data set than the one used in other studies, but similar in length.\footnote{\citet{kouwenberg_markiewicz_verhoeks_zwinkels_2017}, consider the data beginning from 1973. However, data samples from the 1970s and 1980s can vary between data providers and the available quotes are of lower quality than the newer data. For example,
due to illiquidity  of  financial instruments. The details concerning the data sources and any proxies used are presented in the data Appendix \ref{a:data}.}
We use the macroeconomic fundamentals:
\begin{itemize}
\item Uncovered Interest Rate Parity (UIP) 
which postulates that, given the spot rate $S_t$, the expected rate of appreciation (or depreciation) is approximately:
\begin{equation}
\frac{\E(S_{t+h} - S_t)}{S_t} = i_t - i^{*}_t,
\end{equation}
where $i_t$ is the domestic and $i^{*}_t$ is the foreign interest rate corresponding to the time horizon $h$ of the return\footnote{In this context we use the 1 month deposit rates. Theoretically, one should use the 1 month rates from the appropriate cross-currency curves. However, we assume that the difference between the deposit rates in two countries provides a good proxy for the interest rate differential.}.
\item Long-short interest rate difference - the difference between 10 year benchmark government yield and 1 month deposit rate. 
\item Stock growth - monthly return on the main stock index of each of the G10 currencies/countries.
\item Gold price - monthly change in the gold price.
\end{itemize}
The data is standardised based on the mean and standard deviation calibrated to an initial training period of 10 years.

We consider using all possible models as our pool (which consists of 2048 models including all possible subsets of the fundamentals). Comparison to 
the N-average method \citep{diebold2021aggregation}, DeCo \citep{billio2013time} and BPS \citep{McAlinn_2019} 
the competing methods is not available for this pool due to computational cost and so we consider a 
 small pool (which consists of the 32 models based on UIP only and time-constant parameters).
 An exhaustive list of model parameter settings is outlined in Appendix \ref{appendix:fx-model-param}.

We compare performance of the competing model averaging techniques using the logarithmic score and
 economic evaluation using Sharpe ratios of a long-short currency portfolio. We find that LDF provides
 superior performance according to the logarithmic score and demonstrate how these differences in scores manifest themselves in an economic evaluation. 

\subsubsection{Small model pool - analysis of scores} 

\begin{figure}[!tb]    
\begin{tabular}{cc}
Small pool & Large pool \\
         \includegraphics[width=0.4\textwidth]{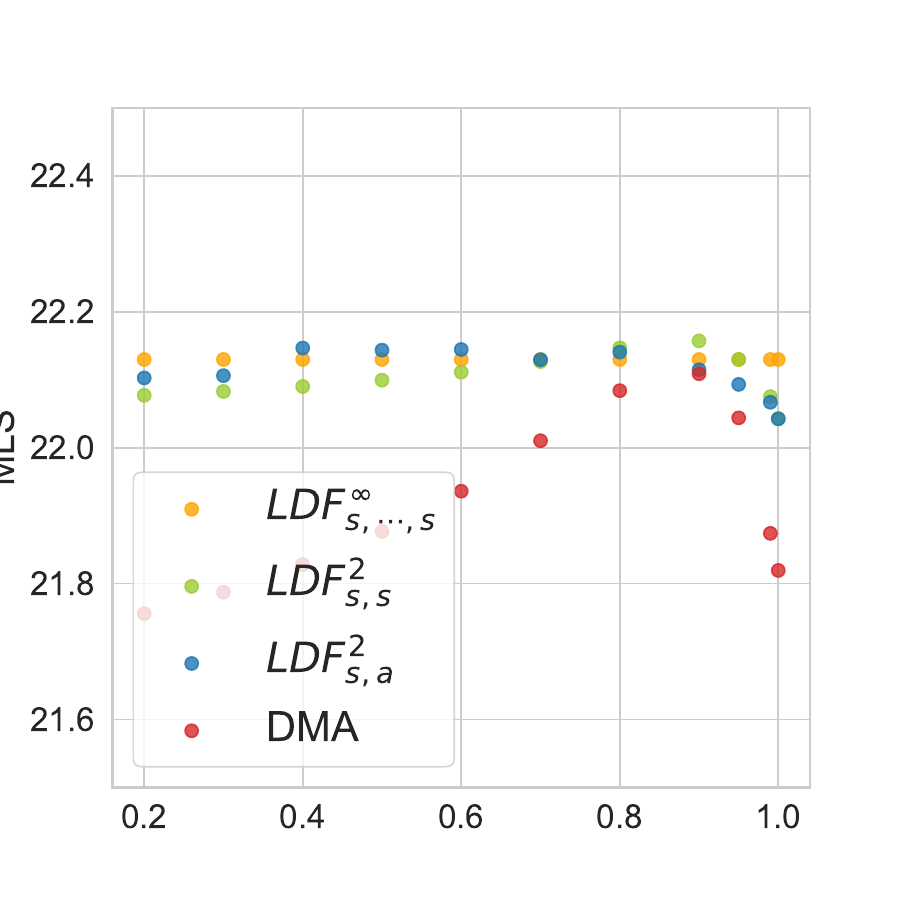}
&
         \includegraphics[width=0.4\textwidth]{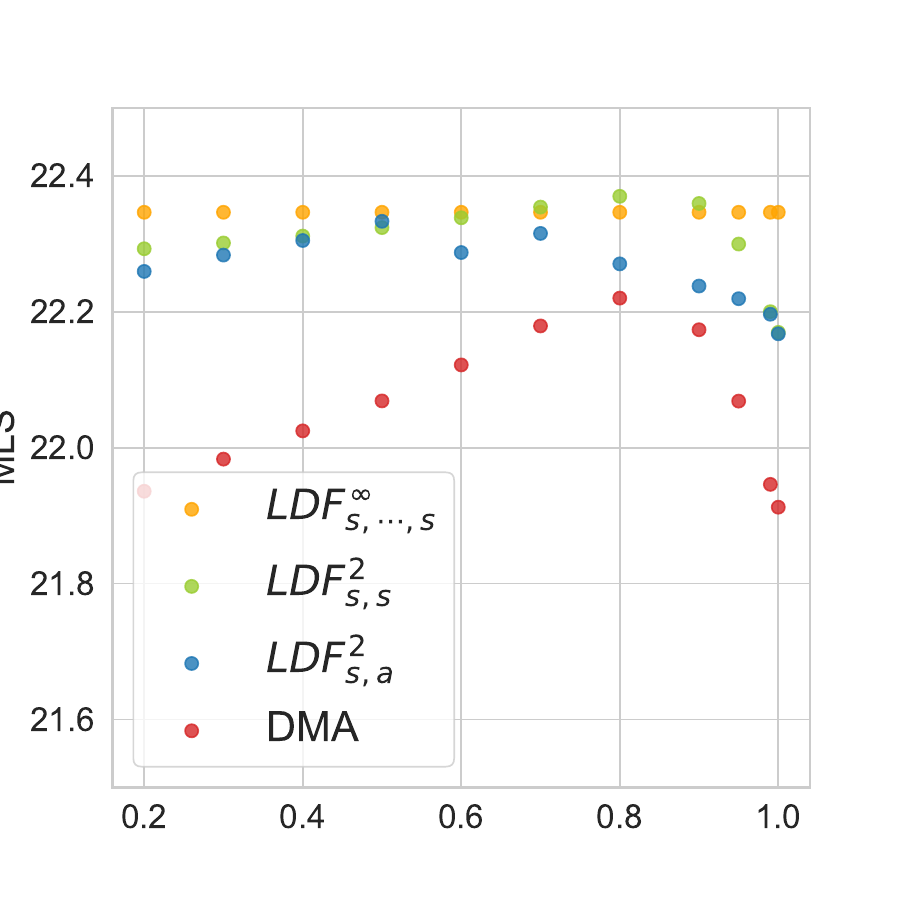}\\
         \includegraphics[width=0.4\textwidth]{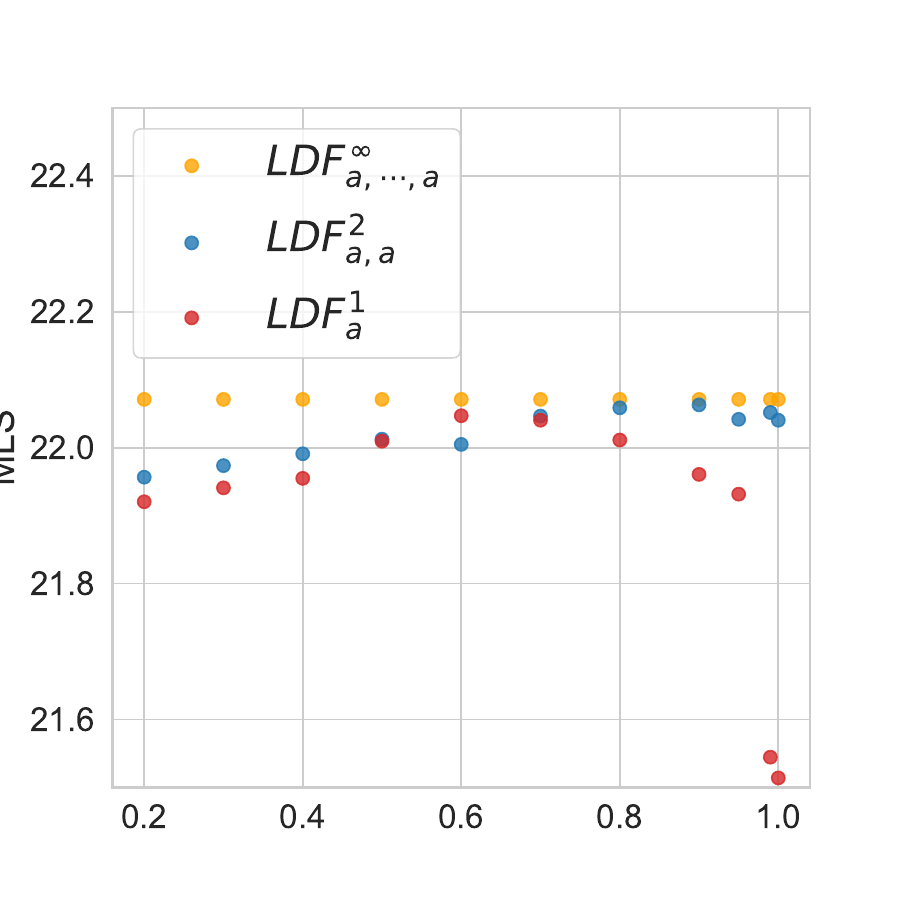}
&
         \includegraphics[width=0.4\textwidth]{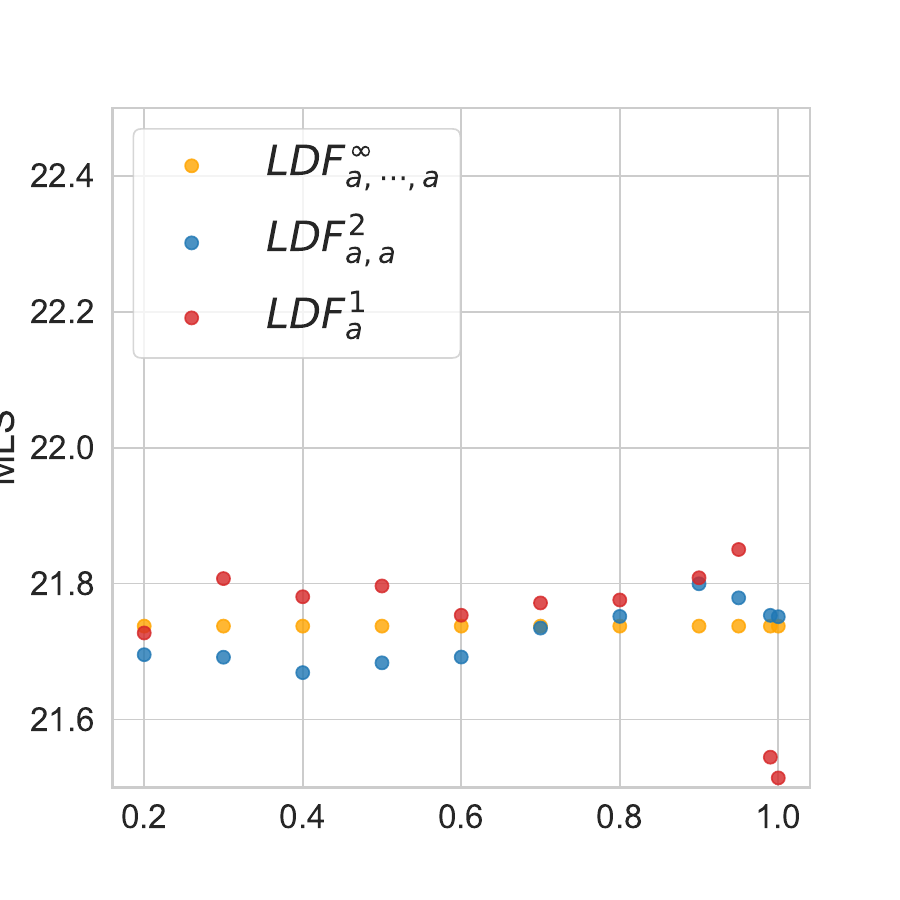}
\end{tabular}
\caption{FX -- MLS versus values of $\alpha$ for LDF and  $\alpha$ for DMA in the x-axis for the small and large model pool. The upper plots show the cases of model averaging whereas the lower plots show model selection.
}
        \label{fig:fxalphadelta}
\end{figure}

 Figure \ref{fig:fxalphadelta} compares the logarithmic score for DMA and some specifications of LDF. 
  LDF provides better performance for an optimal choice of the hyperparamter and is more robust to the choice of the hyperparmeters than  DMA.\footnote{In \ref{appendix:alphadelta} we show that with a dense grid of allowable values for $\alpha$ the points in Figure \ref{fig:fxalphadelta} become smooth curves.}
Interestingly, for model selection, we note that the proposed two-layer LDF specification $\mbox{LDF}_{\mbox{a}, \mbox{a}}^2$ (as well as $\mbox{LDF}_{\mbox{a}, \ldots, \mbox{a}}^{\infty}$) methodology improved upon the DML method \citep{Koop2020}, which as we recall is $\mbox{LDF}_{\mbox{a}, \mbox{a}}^2$ with $\alpha=1$. The best scores in model averaging/selection were achieved for $\mbox{LDF}_{\mbox{s}, \mbox{s}}^2$/$\mbox{LDF}_{\mbox{a}, \mbox{s}}^2$ specification with $\alpha=0.9$. 

The average value of discount parameters $\alpha$ in the first meta-model layer across time, for $\mbox{LDF}_{\mbox{s},\mbox{s}}^2$ with $\alpha=0.9$ is 0.77 which was very similar for $\alpha=1$. 
However, the variability of $\alpha$ in the first meta-model layer for $\alpha < 1$ was much larger, i.e. $\alpha$ being closer to 0 during times of increased volatility and closer to 1 during calmer times (same observation of either pool of models).

\begin{figure}[!tb]    
         \includegraphics[width=1\textwidth]{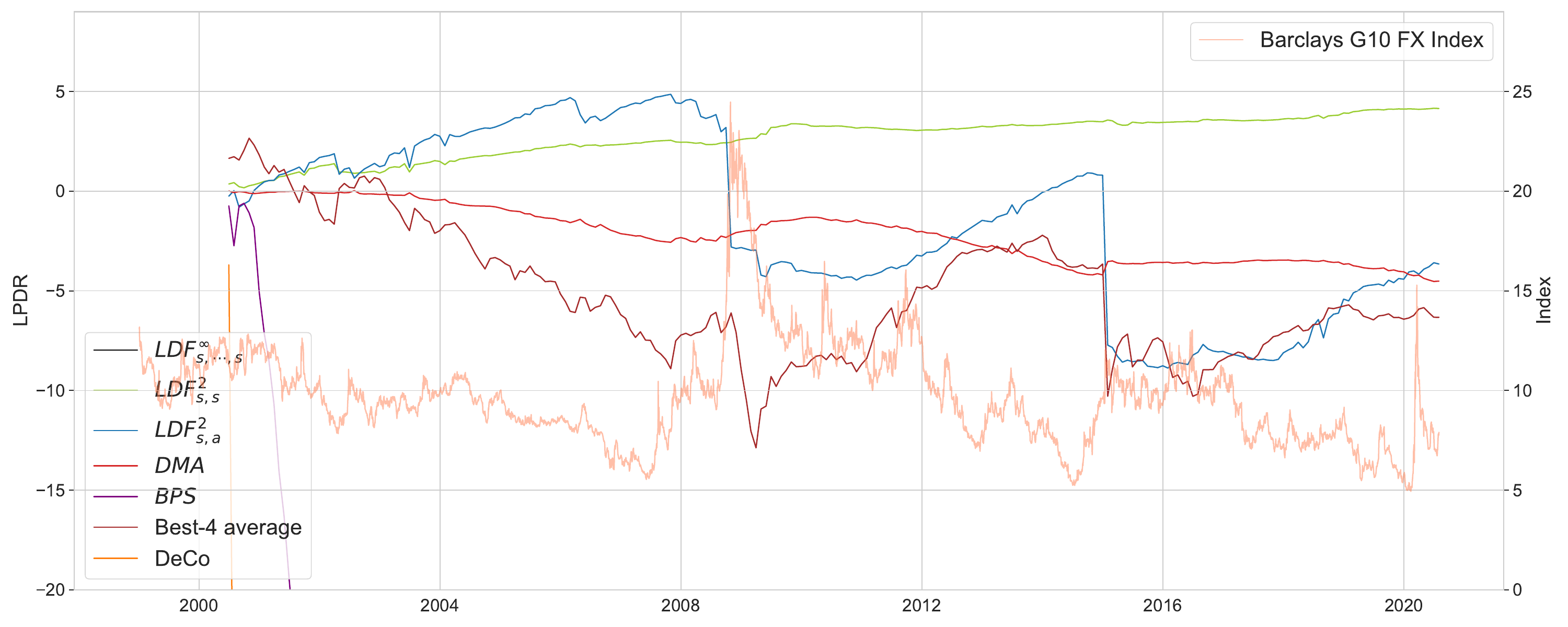}
\caption{FX -- LPDR for model averaging in the small model pool. $\mbox{LDF}_{\mbox{s}, \mbox{s}}^2$ provides best performance robust to increases in the FX volatility.
}
\label{fig:fx_lpdr}
\end{figure}




Figure \ref{fig:fx_lpdr} shows the LDPR for the competing methods on an expanding window with the hyperparameters of LDF calibrated using the first 10 years of data and the competing models calibrated in sample. The LDPRs shows considerable time variation with the sudden drops in performance of $\mbox{LDF}_{\mbox{s}, \mbox{a}}^2$ and Best-4 average models correspond to the period of big FX volatility increases as measured by Barclays G10 FX index.\\
In comparison to other methods,  the $\mbox{LDF}_{\mbox{s}, \mbox{s}}^2$ method with $\alpha=0.9$ performs best (MLS=22.16), followed by other two layer LDF specifications and the 4-model average (MLS=22.10). BPS method (MLS = 21.60) did not perform well here. Similarly, DeCo (MLS = 18.31) method using multivariate normal approximation\footnote{For DeCo, we checked that the marginal distributions are well described by the normal distribution. We then output the covariance matrix from the DeCo source code to complete the multivariate normal approximation.} In terms of model performance out of sample, $\mbox{LDF}_{\mbox{s}, \mbox{s}}^2$ calibrated only on initial 10 years of data (to select $\alpha$) - $\alpha=0.8$ (MLS=22.15) - still outperforms the other non-LDF models which were calibrated in-sample. The detailed results are presented in Table \ref{tab:fx} in \ref{appendix:D}.
The stable state LDF models performed similarly to the two 
layer specification, $\mbox{LDF}_{\mbox{s}, \cdots, \mbox{s}}^{\infty}$ achieves MLS=22.13 and $\mbox{LDF}_{\mbox{a}, \cdots, \mbox{a}}^{\infty}$ scores MLS=22.07. 

\subsubsection{Large model pool - analysis of scores}

We can only consider the LDF methods for the large pool of models due to the run times of the other methods.
Additional meta-model layers have a similar impact as in the small model pool but with 
a less pronounced effect on  model selection and a greater effect on  model averaging Figure~\ref{fig:fxalphadelta}. 
Again, the best model averaging scores were achieved for $\mbox{LDF}_{\mbox{s}, \mbox{s}}^2$ specification with $\alpha=0.8$ (MLS=22.37) and the stable state LDF models performed similarly to the two 
layer specification, 
MLS=22.35 for
$\mbox{LDF}_{\mbox{s}, \cdots, \mbox{s}}^{\infty}$.
For model selection, the multi-layer specification of LDF introduces the robustness to hyperparameters but does not necessarily outperform the single-layer LDF in terms of log-scores. 
Interestingly, in the larger pool, the EWMA Random Walk\footnote{I.e.\, we estimate the volatility of the random walk based on the exponentially weighted moving average.} (RW), decay factor 0.97, model was not the best model of all models considered (MLS = 21.77)  but it performed almost on par with the {\it a posteriori} best model (MLS = 21.78) which indicates that even from a big pool of models it is hard to find a single model that outperforms the random walk. 

\subsubsection{Economic evaluation of model selection}

We consider  economic evaluation  by constructing a portfolio of long and short currency positions targeting 10\% annual volatility with 8bps transaction costs. We measure the performance by looking at the cumulative wealth over time as well as the Sharpe ratio, which captures the risk adjusted performance, applied to 
the smaller model pool of 32 models. To target the Sharpe ratio,  we define the score to be the portfolio returns divided by the the portfolio standard deviation based on a rolling twelve months window. 

\begin{figure}[!tb]    
\begin{tabular}{cc}
         \includegraphics[width=0.4\textwidth]{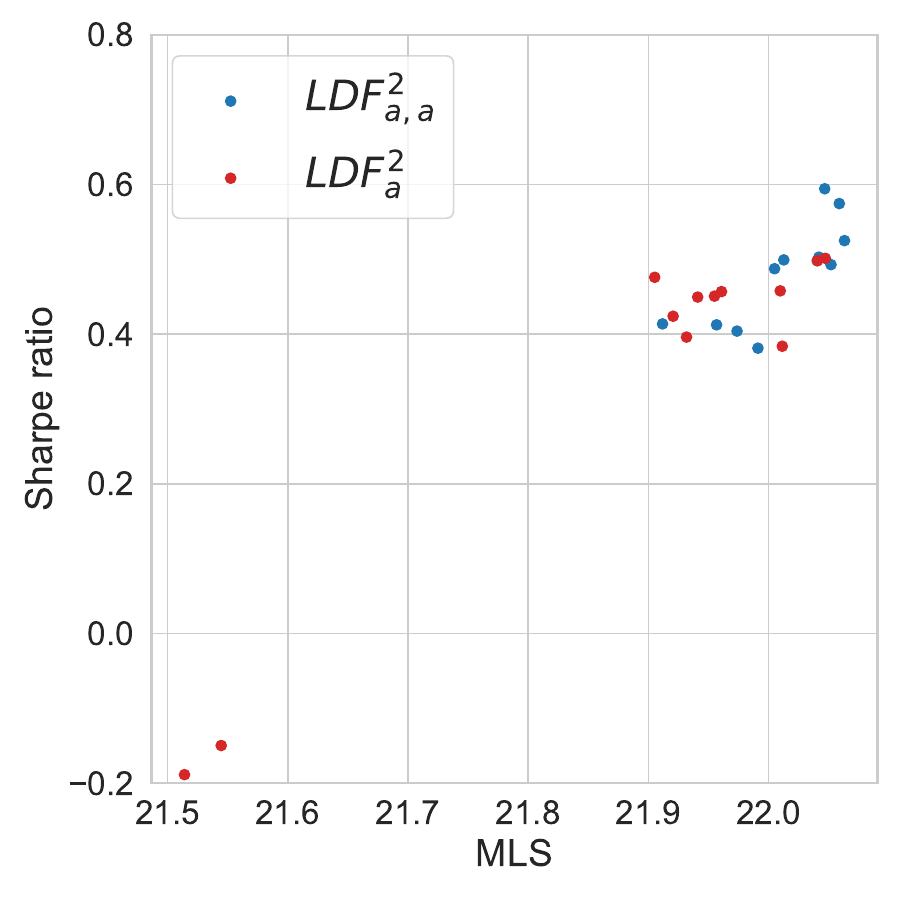}
&
         \includegraphics[width=0.4\textwidth]{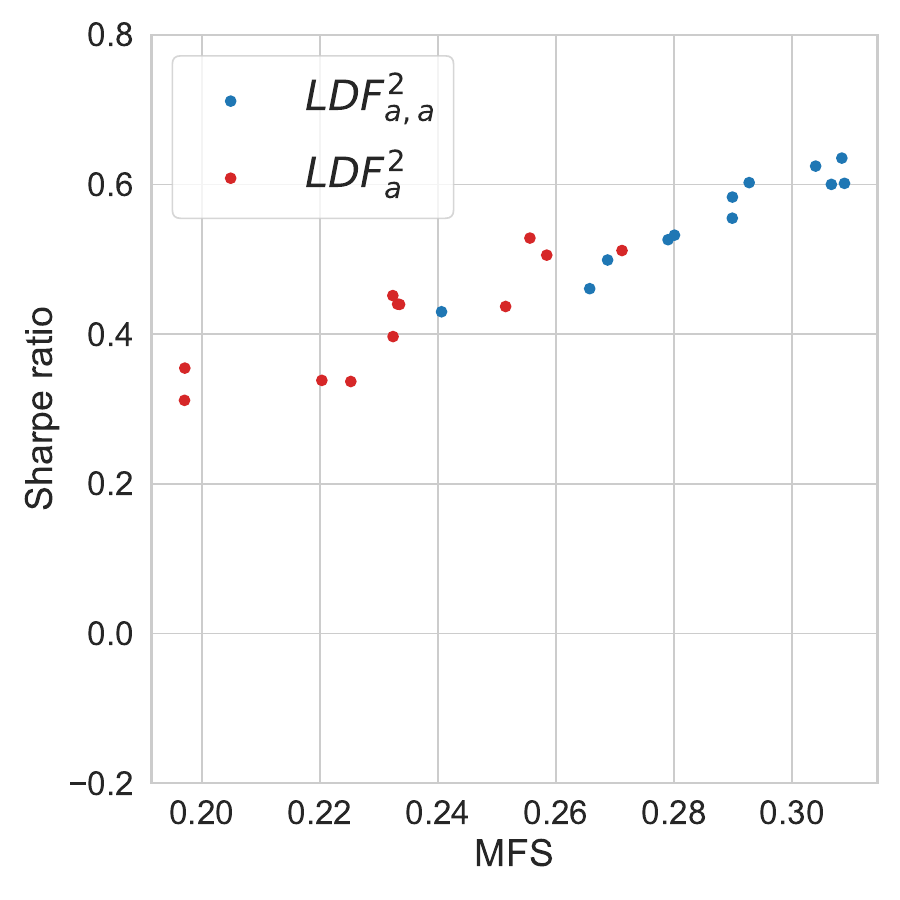}
\end{tabular}
\caption{FX -- mean score values versus achieved Sharpe ratios. In the left had side plot the log scores were used, in the right hand side plot the focused scores were used.
}
    \label{fig:sharpe}
\end{figure}

 We  concentrate on LDF configurations that select a single model at a time, that is $\mbox{LDF}_{\mbox{a}, \mbox{a}}^2$ and $\mbox{LDF}_{\mbox{a}}^1$. Portfolios are constructed by maximizing the returns subject to a fixed risk per model, as in \cite{Koop2020}. Model averaging cannot be used as the correlation between the investment strategies would inevitably change the target risk level of the portfolio. An alternative approach to portfolio construction was presented in \cite{tallman2022bayesian} who use the multivariate focused prediction score in the context of model averaging where each model aims to minimise the risk subject to a fixed return target.
 
  $\mbox{LDF}_{\mbox{a}, \mbox{a}}^2$ only narrowly outperforms $\mbox{LDF}_{\mbox{a}}^1$ on the  log score 
 (Figure \ref{fig:fxalphadelta})  but has a higher Sharpe ratio (Figure \ref{fig:sharpe}). This  is in line with the observations in \citet{Koop2020} who note that small differences in the log scores can translate to noteworthy economic differences.
\begin{figure}[!tb]    
         \includegraphics[width=1\textwidth]{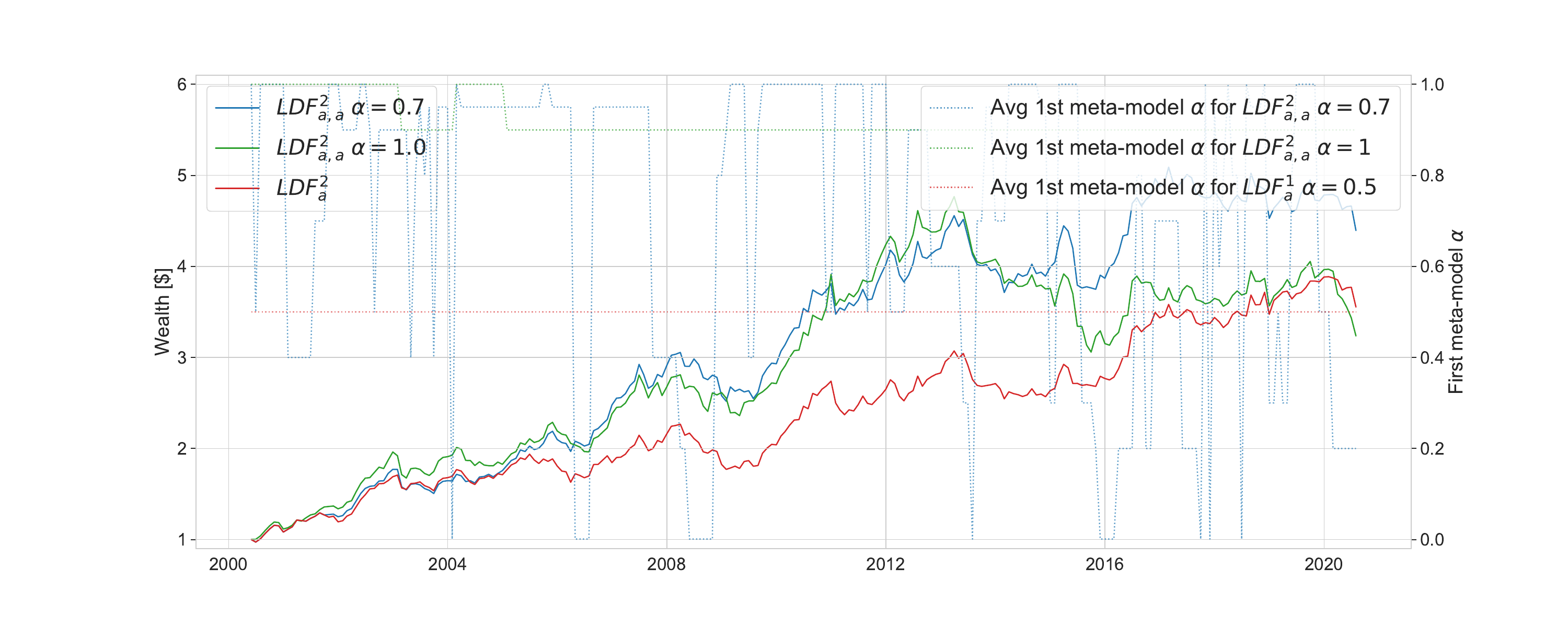}
\caption{FX -- Money through time and discount factor $\alpha$ through time for $\mbox{LDF}_{\mbox{a}, \mbox{a}}^2$ with $\alpha=0.7$ and $\alpha=1$, and $\mbox{LDF}_{\mbox{a}}^1$ with $\alpha=0.5$. 
}
\label{fig:money_focused}
\end{figure}
The right panel of Figure \ref{fig:sharpe} shows the mean focused scores (MFS) where there are clear differences (unlike the log scores) with the double discounting version of LDF achieving better scores leading into higher Sharpe ratios and higher final wealth as seen in Figure \ref{fig:money_focused}. The double discounting of $\mbox{LDF}_{\mbox{a}, \mbox{a}}^2$ allows the discount factor to drop in times of higher volatility such as during the great financial crisis, the Chinese crash or the Brexit referendum. For $\alpha=0.7$ in the second layer the time average value of $\alpha$ in the first layer is 0.71 and with $\alpha=1$ in the first layer it is 0.80. This is in contrast to DML ($\mbox{LDF}_{\mbox{a}, \mbox{a}}^2$ $\alpha=1$) and $\mbox{LDF}_{\mbox{a}}^1$ specifications where in the former the discount factor settles at $0.9$ and does not move and in the latter it is just fixed to a predetermined constant value.

\begin{figure}[!tb]    
         \includegraphics[width=1\textwidth]{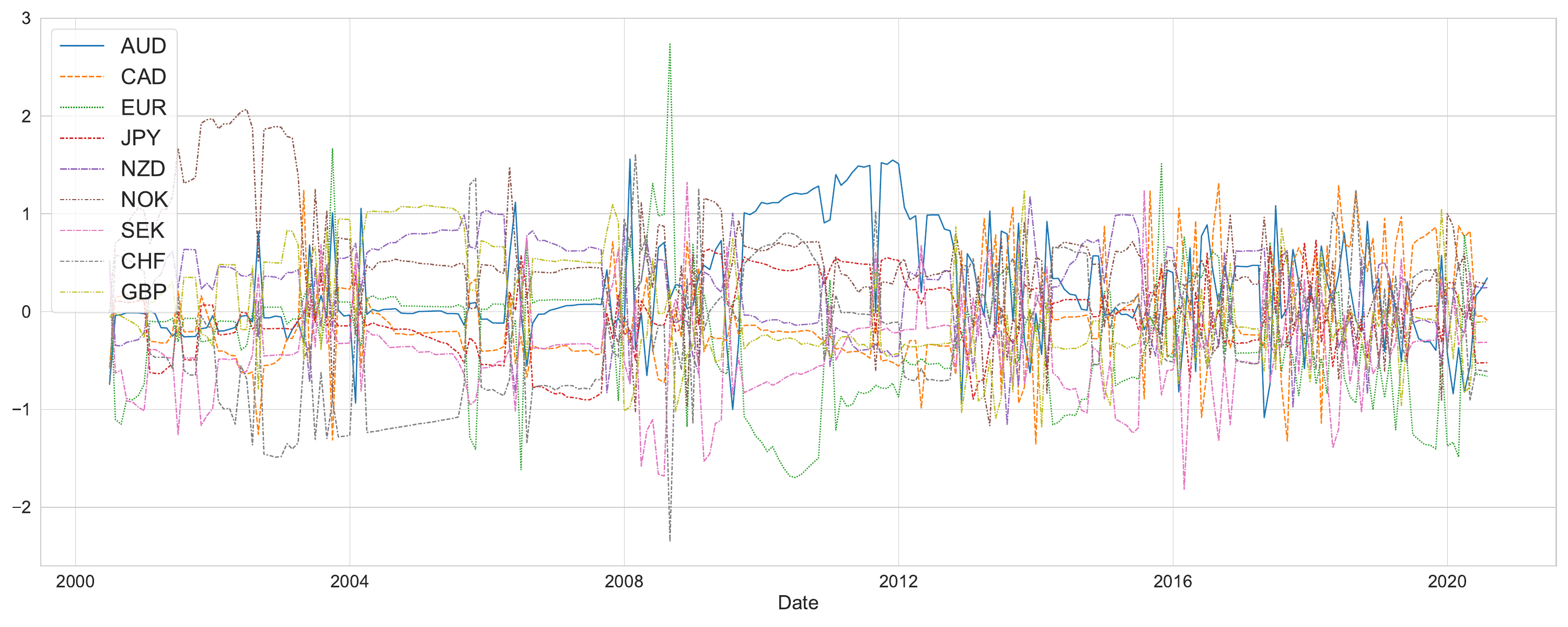}
\caption{FX -- Portfolio composition through time for $\mbox{LDF}_{\mbox{a}, \mbox{a}}^2$ with $\alpha=0.7$. We can clearly see that there are long stretches of stable composition which correspond to the growth periods and the periods of sudden portfolio changes correspond to the times of money growth plateau. 
}
\label{fig:portfolio_focused}
\end{figure}

In Figure \ref{fig:portfolio_focused} we show the portfolio composition though the time. We note that the weights display stability in the times when the portfolio value experiences periods of growth and the sudden weight changes correspond to periods of growth plateau. The weights generally follow the carry trade strategy which is well documented in the literature, see \cite{DellaCorte2012} and references therein.


%% file: parts/subparts/usinf.tex
\subsection{US Inflation Forecasts}
The final study considers an example of \citet{McAlinn_2019}, which
involves forecasting the quarterly US inflation rate between 1961/Q1 and 2014/Q4. Here, the inflation rate corresponds to the annual percentage change in a chain-weighted GDP price index.  There are four competing models:
 $M_1$ includes one period lagged inflation rate, $M_2$ includes period one, two and three lagged inflation interest and unemployment rates, $M_3$ includes period one, two and three lagged inflation rate only and $M_4$ includes period one lagged inflation interest and unemployment rates. All four models provide Student-t distributed forecasts with around 20 degrees of freedom.


The distinguishing features of this example are the small number of models and the existence of 
time periods when none of the models or model combinations lying on simplex provide an accurate mean forecast. In this example we will see the limitation of the LDF and other simplex based methodologies which are unable to correct for forecasting biases if bias corrected models are not explicitly available in the pool. 

The BPS method (MLS = 0.06) dominates all other methodologies since it allows for model combinations not adhering to simplex. In fact, there were six dates in the evaluation period where the mean of BPS synthesised model was greater than the maximum of the underlying models. The feature to go beyond simplex proved to be one of the key factors in the superior performance. 

The next most effective method was N-model average of \citet{diebold2021aggregation} which for $N=2$ and $N=3$ models had a MLS equal to -0.01 and 
provided better performance than the best single model ($M_2$, MLS = -0.02). For $N=2$, out of the 100 evaluation points, the algorithm selected the pair $(M_0, M_1)$ 35 times, the pair $(M_2, M_3)$ 49 times and the pair $(M_1, M_3)$ 16 times. On the other hand, both 2-level LDF model averaging and DMA methods did not work very well in this example but improved upon picking just a single model. The poor performance of 2-level LDF and DMA could mostly be attributed to the highly dynamic nature of these methods which sometimes attached too much weight to a single model that would score poorly.

\begin{figure}[H]
    \centering
    \includegraphics[width=0.4\textwidth, angle=0]{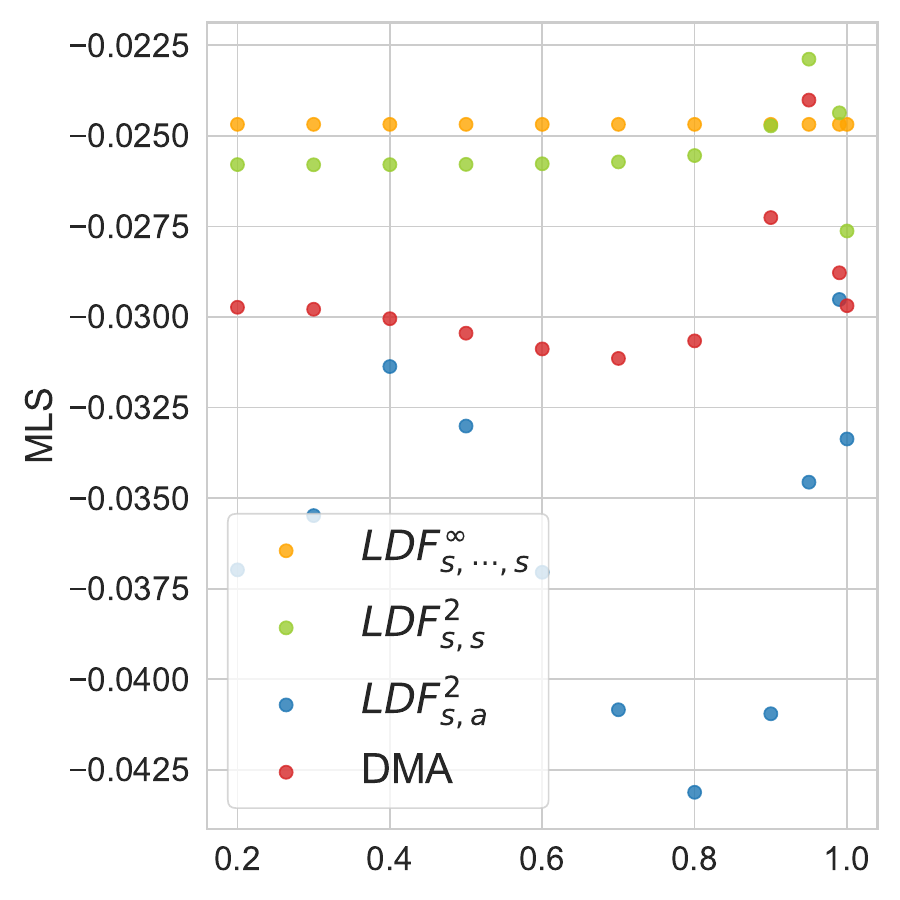}
    \caption{US inflation -- The MLS 
    versus values of $\alpha$ for LDF and $\alpha$ for DML in the x-axis. 
    }
    \label{fig:alphadelta_usinf}
\end{figure}

%% file: parts/discussion.tex
\section{Discussion}             
\label{section:discussion}    

This paper contributes to the model averaging and selection literature by introducing a Loss Discounting Framework which encompasses Dynamic Model Averaging, first presented by \citet{raftery2010online}, generalises Dynamic Model Learning \citep{Koop2020} and introduces additional model averaging or selection specifications. The framework allows for general dynamics for model weights, and works well with focused scores for goal-oriented decision making. The methodology offers extra flexibility which can lead to better forecast scores and yield results which are less sensitive to the choice of hyperparameters. This is particularly important in a more realistic online forecasting setting where selection of the globally optimal hyperparameters is often unattainable.  It also empowers users to choose the model specification in terms of number of levels of discounting layers which is suitable for the problem at hand.

We show that our proposed methodology performs well in both the simulation study as well as in the  empirical examples based on the exchange rate forecasts where we show the superiority of our approach both for model averaging as well as model selection, where for the latter we also demonstrate how the differences in the scores translate to noteworthy economic gains. We find that the LDF can be a good choice when:
 the number of forecasters is fairly large and sophisticated methods become burdensome;
 if we want to have only a small number of hyperparameters to calibrate;
 we suspect that we are in the $\mathcal{M}$-complete/open setting and different models might be optimal at different times but there is no consistent bias to be eliminated across all models;
 if we believe that scoring forecasters on the joint predictive density or joint utility basis is reasonable.

The LDF is by no means the panacea for model synthesis and the performance of different model synthesis methods depends on the problem (as seen in the empirical studies). However, LDF is often able to achieve competitive performance with a low computational overhead by using flexible dynamics and general model scores in an  easy-to-implement and compute framework.




There are multiple open avenues to explore.
Many current forecast combination methods described in the literature assume that the pool of forecasters does not change over time
\citep[see {\it e.g.}][]{raftery2010online, diebold2021aggregation, McAlinn_2019}. In some situations this is a substantial limitation, for example, if the forecasts are provided by a pool of experts.

Let us first consider the situation of a new agent being added to the existing pool of forecasters. The existing forecasters already have a track record of forecasts and corresponding scores. A new forecaster could be included with an initial weight. This could be fairly easily achieved in the LDF by considering a few initial scores. It is not clear what this weight should be, especially in more formal methodologies which relax the simplex restriction like \citet{McAlinn_2019}. Similarly, forecasters may drop out completely or for some quarters before providing new forecasts. Again, in general, it is hard to know how to weight these forecasters. The LDF provides a rationale, we should be using an estimate of that forecaster's score when a forecast is made. This is a time series prediction problem and can be approached using  standard methods.

We noted in the empirical sections that the best performing discount factor in the second layer is larger than the average across time discount factor in the first layer. We showed that as one keeps adding more and more layers of meta-models the weights converge to an equilibrium. I.e. adding more layers does not change the scores any more and any choice of the discount factor in the final layer leads to the same score and same discount factors in all other layers. 

It would also be of interest to consider the case when the models to be combined themselves allow for sharp breaks (\cite{gerlach2000efficient}, \cite{huber2019should}). Intuitively, more flexible models will lead to weights with less fluctuations if the models are able to represent the true DGP (for example, if one model is correctly specified then LDF should be able to roughly replicate BMA). We believe that the use of out-of-sample log predictive scores to calculate weights in LDF will avoid the problems of overfitting found using in-sample estimation methods. Therefore, we believe that LDF can take advantage of more flexible models and robustifies against the use of overly simple models.

As mentioned before, in most examples, we use joint predictive log-likelihood as a statistical measure of out-of-sample forecasting performance. It gives an indication of how likely the realisation of the modelled variable was conditional on the model parameters. The logarithmic scoring rule is strictly proper but it severely penalises low probability events and hence it is sensitive to tail or extreme cases, see \citet{gneiting2007strictly}. A different proper scoring rule could be used when needed or if a decision is to be made based on the outcomes of model averaging/selection then a focused score (or utility), aligned with the final goal, can be used as successfully demonstrated in one of our examples.

Furthermore, since the scoring function is often based on the joint forecast probability density function, our methodology is not best suited to take strength from forecasters who might be good at forecasting one or more variables but not the others. This is partially due to the fact that our methodology does not consider any dependency structure between expert models and the weighting is solely performance based. An extension introducing a way to take the agent inter-dependencies into consideration would be of considerable interest.



More broadly, the exponential discounting recipe could be generalised and expanded by any forecast of the scores which could involve more parameters.

%% file: parts/appendices/technical_appendix.tex
\section{Asymptotic properties of Loss Discounting Framework}
\label{appendix:X}  

\begin{theorem}[Convergence for softmax]
For $\mbox{LDF}^n_{ss\dots s}(S)$ and $\mbox{LDF}^{n+1}_{ss\dots s}(S)$ models, then let $X_1^{(n)}, X_2^{(n)},\ldots \sim p^{(n)}_m$ and $X_1^{(n+1)}, X_2^{(n+1)}\ldots \sim p^{(n+1)}_m$ then $n\to\infty$ $X^{(n)} \xrightarrow[]{d} X^{(n+1)} \xrightarrow[]{d} X$ for any $k=1,\ldots,K$, where $X \sim p^*$ with $p^*$ being some probability distribution function.
\end{theorem}

\begin{proof}
First let us recall that for $n \geq 2$ a model $p^{(n)}_m(y_t|y_{t-1})$ is a weighted average of the models  in the previous layer and each weight $w \in [0,1]$. In effect in each layer we have some average of models\\ $p^{(n)}_m(y_t|y_{t-1}) = \sum_{k=1}^K \omega^{(n)}_{t|t-1,k}(m) p_k^{(0)}$, as shown in Eq.\ \refeq{ldf-recursive}. This implies that 
\begin{equation*}
    \omega^{(n)}_{t|t-1,k}(m) \in \left[\min(\{\omega^{(n-1)}_{t|t-1,k}(m) \}_{m=1}^M), \max(\{\omega^{(n-1)}_{t|t-1,k}(m) \}_{m=1}^M)\right]
\end{equation*}
 which we call $\mathcal{S}_{\omega_{k}}^{(n)}(m)$ for $m=1,\ldots,M$ and $k=1,\ldots,K$. From this can see that $[0,1] \supseteq \mathcal{S}_{\omega_{k}}^{(1)}(m) \supseteq \mathcal{S}_{\omega_{k}}^{(2)}(m) \cdots$ so that $\lim_{n\to\infty} \mu \left( \mathcal{S}_{\omega_{k}}^{(n)}(m) \right) = 0$, where $\mu$ is the Lebesgue measure on $\mathbb{R}$ as long as the weights $w$ are not all $\{ 0,1\}$ which for the softmax function is satisfied almost surely. Since this is true for any $m$ we have that in the limit $n \rightarrow \infty$  ${\omega_{k}}^{(\infty)}(1) = {\omega_{k}}^{(\infty)}(2) = \ldots = {\omega_{k}}^{(\infty)}(M)$ $\forall k$.
\end{proof}

\begin{corollary}[Convergence for argmax]
For $\mbox{LDF}^n_{aa\dots s}(S)$ and $\mbox{LDF}^{n+1}_{aa\dots s}(S)$ models, then let $X_1^{(n)}, X_2^{(n)},\ldots \sim p^{(n)}_k$ and $X_1^{(n+1)}, X_2^{(n+1)}\ldots \sim p^{(n+1)}_k$ then $n\to\infty$ $X^{(n)} \xrightarrow[]{d} X^{(n+1)} \xrightarrow[]{d} X$ for any $k=1,\ldots,K$, where $X \sim p^*$ with $p^*$ being some probability distribution function.
\end{corollary}
\begin{proof}
The proof of this result is harder than the previous one since all weights are always either 0 or 1 and, ultimately, only a single model will be selected. This means that for $\mathcal{S}_{p^{(1)}}=\{ p_k^{(1)}(y_t|y_{t-1}) \}_{m=1}^{M}$ we can have at most $\min(K,M)$ distinct models. At each next layer we will have $\min(K,M)$  or less distinct models so that $\mathcal{S}_{p^{(1)}} \supseteq \mathcal{S}_{p^{(2)}} \cdots$. Therefore, $\lim_{n\to\infty} \left| \mathcal{S}_{p^{(n)}} \right| =1 $ almost surely.
\end{proof}

%% file: parts/appendices/appendix_simulation.tex
\section{Simulation study - supplementary material}
\label{appendix:A}    
In this appendix we provide additional details corresponding to the simulation example from section \ref{section:simulation} as well as the results of additional experiments performed in order to check the robustness and persistence of the results presented.
\begin{itemize}
\item time constant Markov switching levels
\item time varying Markov switching levels
\end{itemize}

\subsection{Simulation study - additional results}
In Table \ref{tab:table1} we present the full set of results from the simulation study in section \ref{section:simulation}. Apart from MSE we also provide the sums of log-scores which also show that the two-level LDF models also dominate in this performance metric.
\begin{table}
\centering
\resizebox{\textwidth}{!}{%
\input{tables/table_simulation.tex}
}
\caption{Predictive log-likelihood for our simulated example averaged over $R=10$ runs and the associated standard deviation. We denote the average of the mean log score as $\bar{\mbox{MLS}} = \frac{1}{TR} \sum\limits_{r} \sum\limits_{t} \log(p)$, the average of the cumulative log score as $\overline{\sum \log(p)} = \frac{1}{R} \sum\limits_{r} \sum\limits_{t} \log(p)$ and by $\sigma(.)$ the corresponding standard deviation of the quantities in question. We can see that our proposed model outperforms all other methods.}
\label{tab:table1}
\end{table}

\subsection{Simulation study - BPS configuration and extended commentary}
The Bayesian Predictive synthesis model was run with the following sets of configurations and corresponding outcomes. Each run was performed using 8000 Monte Carlo draws from which 3000 were used for the burnin. We used parallel computing with 34 processes on Intel(R) Xeon(R) Gold 6140 CPU @ 2.30GHz. Each BPS run took around 42 hours. 
\begin{itemize}
    \item $\beta = 0.95$, $\delta = 0.95$, $C_0 = \text{diag}(0.0021, \ldots, 0.0021)$, $s_0=0.09$: resulting in MLS = -0.73
    \item $\beta = 0.95$, $\delta = 0.95$, $C_0 = \text{diag}(0.0021, \ldots, 0.0021)$, $s_0=0.12$: resulting in MLS = -0.73
    \item $\beta = 0.95$, $\delta = 0.95$, $C_0 = \text{diag}(21.0, \ldots, 21.0)$, $s_0=0.09$: resulting in MLS = -0.74 
    \item $\beta = 0.99$, $\delta = 0.99$, $C_0 = \text{diag}(0.0021, \ldots, 0.0021)$, $s_0=0.09$: resulting in MLS = -0.95
\end{itemize}
One of the two major reasons why BPS performed worse than LDF in the simulation study is that BPS produces synthesised forecasts which are maginally normally distributed. Whereas, LDF is a mixture of normals and hence fatter tailed. Furthermore the root mean square error of the $\text{LDF}^{\infty}_{s,\ldots,s}$ is 0.66 versus for BPS 0.76.

\subsection{Time-constant Markov switching model}
In this experiment we adopt the same set up as in Section \ref{section:simulation} but we set the Markov transition matrix for $\mu_t$ to 
$Q=\begin{pmatrix}
0.990 & 0.005 & 0.005\\
0.005 & 0.990 & 0.005\\
0.005 & 0.005 & 0.990\\
\end{pmatrix}$ for three states $\{-1, 0 ,1\}$ and the rest of parameters we set to the same values as in the Section \ref{section:simulation}, namely, $\phi_x=0.9$, $\sigma_x = 0.3$, $\sigma_y = 0.3$, $\sigma_{tk} = 0.1 \, \forall k$, $K=20$, $T=2001$. We compare our 2-level method to the plain DMA of \citet{raftery2010online}. We run the experiment 10 times. The results are reported in first two columns from the left in Table \ref{tab:table3}. Interestingly our method outperforms the standard DMA for all $\alpha$ parameters reported here which suggest a good degree of robustness of the method.

\begin{table}[!tb]
\centering
\resizebox{\textwidth}{!}{%
\begin{tabular}{c|cccc||cccc}
                      &\multicolumn{4}{c||}{\textbf{Constant transition matrix}} & \multicolumn{4}{c}{\textbf{Time-varying transition matrix}}             
\\ \hline \hline
\textbf{Model}        & $\bar{\mbox{MLS}}$      & $\sigma[\mbox{MLS}]$ &  $\overline{\sum log(p)}$ &  $\sigma[\sum \log(p)]$ & $\bar{\mbox{MLS}}$      & $\sigma[\mbox{MLS}]]$   & $\overline{\sum log(p)}$ & $\sigma[\sum \log(p)]$ \\
\hline
\hline
\multicolumn{9}{c|}{\textbf{Dynamic Model Averaging}}               \\
\hline
$\alpha=0.95$              & -0.59             & 0.08   & -1170.88 & 154.05 & -0.73    & 0.10 & -1450.78 & 192.49 \\
$\alpha=0.90$              & -0.53             & 0.07   & -1057.20 & 137.32 & -0.64    & 0.09 & -1272.78 & 171.20 \\
$\alpha=0.80$              & -0.47             & 0.05   & -935.10  & 105.33 & -0.54    & 0.07 & -1069.35 & 131.22 \\
$\alpha=0.70$              & -0.45             & 0.04   & -895.58  & 81.33  & -0.50    & 0.05 & -988.16  & 100.89 \\
$\alpha=0.60$              & -0.45             & 0.03   & -898.27  & 61.24  & -0.49    & 0.04 & -964.58	 & 76.44  \\
\hline \hline
\multicolumn{9}{c|}{$\mbox{LDF}_{\mbox{s}, \mbox{a}}^2$} \\
\hline
$\alpha =1.00$             & -0.45             & 0.04   & -899.77  & 72.40  & -0.49    & 0.04 & -973.77  & 75.43 \\
$\alpha =0.95$             & -0.43             & 0.04   & -860.64  & 80.33  & -0.47    & 0.04 & -938.28  & 80.18 \\
$\alpha =0.90$             & -0.44             & 0.04   & -866.64  & 79.01  & -0.48    & 0.04 & -949.11  & 78.41 \\
$\alpha =0.80$             & -0.44             & 0.04   & -875.42  & 79.34  & -0.49    & 0.04 & -960.08	 & 80.10 \\
\hline \hline
\multicolumn{9}{c|}{$\mbox{LDF}_{\mbox{s}, \mbox{s}}^2$} \\
\hline
$\alpha =1.00$             & -0.45             & 0.04   & -891.39  & 71.39  & -0.49    & 0.04 & -965.73  & 75.61 \\
$\alpha =0.95$             & -0.40             & 0.03   & -797.36  & 63.64  & -0.44    & 0.04 & -874.81  & 69.77 \\
$\alpha =0.90$             & -0.39             & 0.03   & -778.02  & 57.32  & -0.43    & 0.03 & -849.84	 & 64.98 \\
$\alpha =0.80$             & -0.39             & 0.03   & -774.92  & 50.41  & -0.42    & 0.03 & -838.88	 & 58.37 
\end{tabular}%
}
\caption{Predictive log-likelihoods for our additional simulation study averaged over 10 runs. We can see that LDF based model outperforms the standard DMA and static online learning.}
\label{tab:table3}
\end{table}

\subsection{Time-varying Markov switching model}
In this simulation study we changed the definition of the transition matrix to be:
$Q_t=\begin{pmatrix}
0.990 & 0.005 & 0.005\\
0.005 & 0.990 & 0.005\\
0.005 & 0.005 & 0.990\\
\end{pmatrix}$ for $t<1000$ and $Q_t=\begin{pmatrix}
0.980 & 0.010 & 0.010\\
0.010 & 0.980 & 0.010\\
0.010 & 0.010 & 0.980\\
\end{pmatrix}$ for $t>=1000$. The rest of parameters remains as defined before. The results are reported in first two columns from the right in Table \ref{tab:table3}. In this case, similarly to the time-constant transition matrix example, our model shows a superior performance over the benchmarks.

%% file: tables/table_simulation.tex
\begin{tabular}{c|c|c|c|c||c|c|c|c|c}
\textbf{Model}    & $\bar{\mbox{MLS}}$   & $\sigma[\mbox{MLS}]$ & $\overline{\sum log(p)}$ &$\sigma[\sum log(p)]$& \textbf{Model}  & $\bar{\mbox{MLS}}$   & $\sigma[\mbox{MLS}]$ & $\overline{\sum log(p)}$ &$\sigma[\sum log(p)]$  \\
\hline
\hline
BMA               & -4.34           & 0.05               & -8601.11	      & 107.41	        & BPS               & -0.73           &$\text{N/A}^*$& -1444.08 &$\text{N/A}^*$\\
\hline 
\multicolumn{5}{c||}{\textbf{Best N-average, rolling-window=5}}                                                 &  \multicolumn{5}{c}{$LDF_{\mbox{s}, \mbox{a}}^2$}\\
\hline
N=1               & -0.71           & 0.02               & -1404.45        & 45.73           & $\alpha =1.00$   & -0.50           & 0.02      & -1000.88    &  35.00 \\
N=2               & -0.55           & 0.03               & -1080.86	       & 52.25	         & $\alpha =0.95$   & -0.46           & 0.02      & -919.52     &  45.95 \\
N=3               & -0.52           & 0.03               & -1031.68	       & 49.74           & $\alpha =0.90$   & -0.47           & 0.03      & -932.42     &  52.32 \\
N=4               & -0.52           & 0.02               & -1033.21        & 33.46           & $\alpha =0.80$   & -0.48           & 0.03      & -955.35     &  56.16 \\
N=5               & -0.54           & 0.02               & -1063.02        & 31.84	         & $\alpha =0.70$   & -0.49           & 0.03      & -966.49     &  57.96 \\   
N=6               & -0.57           & 0.01               & -1134.41	       & 27.37           & $\alpha =0.60$   & -0.49           & 0.04      & -978.41     &  58.88 \\   \cline{6-10}
\hline
\multicolumn{5}{c}{$LDF_{\mbox{s}, \mbox{s}}^2$} & \multicolumn{5}{c}{\textbf{Dynamic Model Averaging}}                               \\
\hline
$\alpha =1.00$    & -0.49          & 0.02                & -973.02        &  34.45          & $\alpha = 1.00$   & -0.80           & 0.03	    & -1585.42  &  50.03 \\
$\alpha =0.95$    & -0.43          & 0.02                & -848.97        &  40.46          & $\alpha = 0.95$   & -0.70           & 0.02	    & -1395.40  &  46.96 \\
$\alpha =0.90$    & -0.42          & 0.02                & -832.41        &  39.40          & $\alpha = 0.90$   & -0.63           & 0.02	    & -1237.18  &  49.68 \\
$\alpha =0.80$    & -0.42          & 0.02                & -822.47        &  35.85          & $\alpha = 0.80$   & -0.54           & 0.02	    & -1063.35  &  43.70 \\
$\alpha =0.80$    & -0.42          & 0.02                & -824.96        &  35.05	        & $\alpha = 0.70$   & -0.50           & 0.02        & -993.32	&  38.20 \\
$\alpha =0.60$    & -0.42          & 0.02                & -830.30        &  34.40          & $\alpha = 0.60$   & -0.49           & 0.02	    & -970.13	&  34.55                  
\end{tabular}%

%% file: parts/appendices/appendix_param_c.tex
\section{Parameter c}
\label{appendix:B}  
As described in detail in \citet{Yusupova2019} the parameter $c$ introduced by \citet{raftery2010online} to ``avoid a model probability being brought to machine zero by aberrant observations'' can have a sizeable impact on the model averaging algorithm performance. This parameter gives a small extra weight to all models which means that the final model density combination has fatter tails. 

This can beneficial when we are dealing with a regime switching Gaussian process. In this situation our model averaging will necessarily lag behind the observations (since it will take a few observations to realise that regime switch occurred) which in turn will cause a number of predictive posterior distributions to miss the realisations by a potentially large margin. If the model is scored using logarithmic rule the penalty for such misses will be quite high. However, a fatter-tailed predictive posterior distribution will incur less severe losses in this case.

However, if the regime switches are infrequent the fatter-tailed predictive posterior distribution will on average receive a lower log-score as opposed to a distribution mixture with thinner tails.

In our research we set the parameter $c=10^{-20}$, just to avoid the machine zero probabilities of models but to avoid introducing fatter-tailed predictive distributions. This was done to aim for a fair comparison between various model averaging algorithms. The tuning of this parameter is outside of the scope of this paper. 

%
%

%% file: parts/appendices/appendix_fx.tex
\section{Foreign Exchange Study - supplementary materials}
\label{appendix:D}

\subsection{TVP VAR model overview}
\label{appendix:tvp-var}
Vector Autoregressive (VAR) model is a generalisation of a univariate autoregressive model to multiple variables. All variables enter the model the same way and each variable has an equation involving its own lags, cross-lags and, potentially, exogenous variables. The time-varying version allows the parameters of the vector autoregression to be functions of time. The TVP-VAR model used for this research is as specified in \citet{Koop2020}:
\begin{align}
\label{eq:var}
	y_t &= X_t \beta_t + \epsilon_t, \quad \epsilon_t \sim N(0,\Sigma_t), \\
\label{eq:var2}
	\beta_{t+1} &= \beta_t + u_t, \quad u_t \sim N(0,Q_t),
\end{align}
where $y_t$ is an $m \times 1$ vector containing observations of $m$ variables (in this case, discrete exchange rate returns), $X_t$ is a model matrix where each row contains variables of a single VAR equation. That is, $X_t$ contains: intercept, $p$ lags of endogenous variables (own lags and cross-lags) and a single lag of exogenous variables (economic factors). The set of exogenous variables is divided into two classes: asset specific, with $n_x$ elements and non-asset specific, with $n_{xx}$ variables. Therefore, the coefficient vector $\beta_t$ has $k=m(1+p \cdot m + n_x + n_{xx})$ elements.

We utilise a variant of the Minnesota prior:
\begin{equation}
\beta_0 \sim N(0,\Omega_0).
\end{equation}
This setup means that the expected values of the coefficient vector $\beta_0$ are initialised as a vector of zeroes with the covariance matrix $\Omega_0$. The Minnesota prior assumes a diagonal structure of the covariance matrix where the size of the elements determine the strength of shrinkage of the respective coefficients. The smaller the diagonal elements the stronger the shrinkage towards 0

\subsection{TVP VAR model parameters}
\label{appendix:fx-model-param}
The model settings are as follows:
\begin{itemize}
\item \textbf{Discount factor $\kappa$} - we follow \citet{morgan1996reuters} and set $\mathcal{S}_{\kappa}=\{0.97\}$.
\item \textbf{Number of lags $p$} - we follow \citet{Koop2020} and set number of lags to $\mathcal{S}_p=\{6\}$.
\item \textbf{Discount factor $\lambda$} - we verify the findings by \citet{abbate2016point}, suggesting that the time variation of VAR coefficients $\beta_t$ is not desirable from the forecasting performance perspective by comparing the forecast performances between $\mathcal{S}_{\lambda}=\{1\}$ and $\mathcal{S}_{\lambda} = \{0.5, 0.7, 0.9, 1 \}$.
\item \textbf{Discount factor $\alpha$}
\item \textbf{Intercept shrinkage parameter $\gamma_1$} - we consider a grid $\mathcal{S}_{\gamma_1} = \{0,10\}$.
\item \textbf{Endogenous variable shrinkage parameters $\gamma_2,\gamma_3$} - the models are based on the grid of possible values $\mathcal{S}_{\gamma_2} = \mathcal{S}_{\gamma_3} = \{0,0.1,0.5,0.9\}$. 
\item \textbf{Exogenous variable shrinkage parameters $\gamma_4, \ldots, \gamma_{n_x + n_{xx} + 3}$} - we consider a binary grid $\mathcal{S}_{\gamma_4} = \ldots = \mathcal{S}_{\gamma_{n_x + n_{xx} + 3}} = \{0,1\}$, i.e., a variable is either included or not.
\item \textbf{Target volatility} - we follow \citet{Koop2020} and \citet{DellaCorte2012} and set the target annual volatility to $\sigma=0.1$ (10\%).
\item \textbf{Transaction costs} - we follow \citet{Koop2020} and set the transaction costs to be the fixed, symmetric value $\tau=0.0008$ (8bps).
\end{itemize}

To evaluate the models we first specify the allowable BVAR model space/universe by defining the model parameter restrictions, data sample, regressors, and covariance matrix estimator. Those definitions translate into a set of possible models that an investor can choose at any point in time by taking all possible permutations of the model parameter restrictions. For example, if we restrict the economic variables to the UIP only and impose further restrictions on $\lambda$ and $\alpha$, i.e. $\mathcal{S}_{\lambda}=\{1\}$, $\mathcal{S}_{\alpha}=\{1\}$, then a model universe comprises of a set of 64 possible models ($|\mathcal{S}_{\gamma_1}| \cdot |\mathcal{S}_{\gamma_2}| \cdot |\mathcal{S}_{\gamma_3}| \cdot |\mathcal{S}_{\gamma_4}| \cdot |\mathcal{S}_{\lambda}| \cdot |\mathcal{S}_{\alpha}| = 2 \cdot 4 \cdot 4 \cdot 2 \cdot 1 \cdot 1 = 64$, where $|\mathcal{S}|$ denotes the number of elements in set $\mathcal{S}$ and $\mathcal{S}_{\gamma_4}$ is the set of shrinkage parameters for the exogenous variable UIP).

\subsection{Dense grid results}
\label{appendix:alphadelta}
\begin{figure}[!tb]    
 \centering
     \begin{subfigure}[b]{0.49\textwidth}
         \centering
         \includegraphics[width=\textwidth]{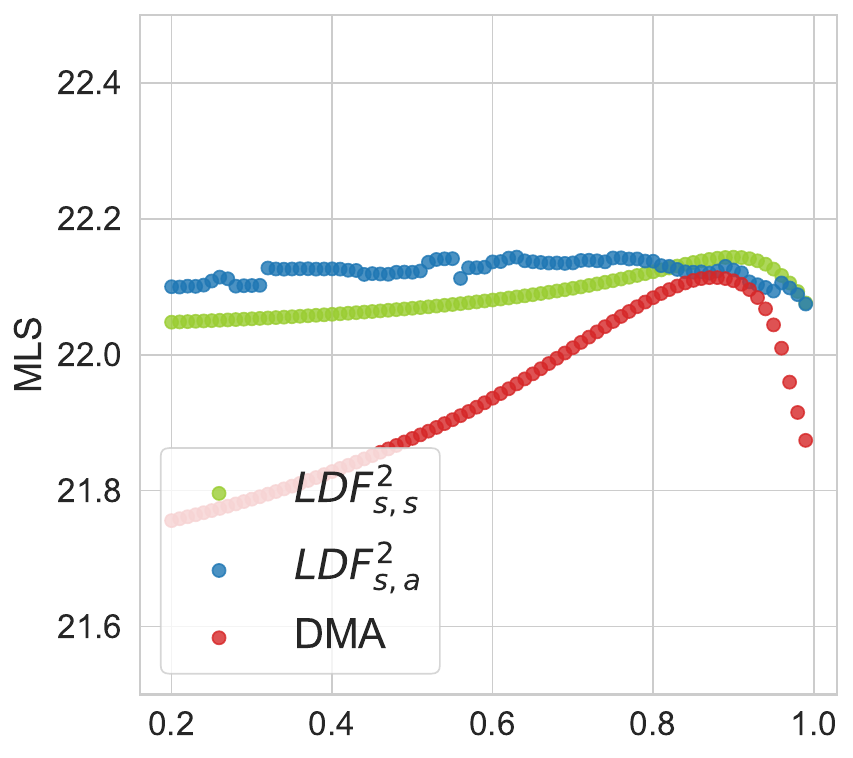}
         \caption{MLS versus values of $\alpha$ for LDF and  for DMA in the x-axis, for the small model pool}
         \label{fig:fxalphadelta_small_dense}
     \end{subfigure}
     \hfill
     \begin{subfigure}[b]{0.49\textwidth}
         \centering
         \includegraphics[width=\textwidth]{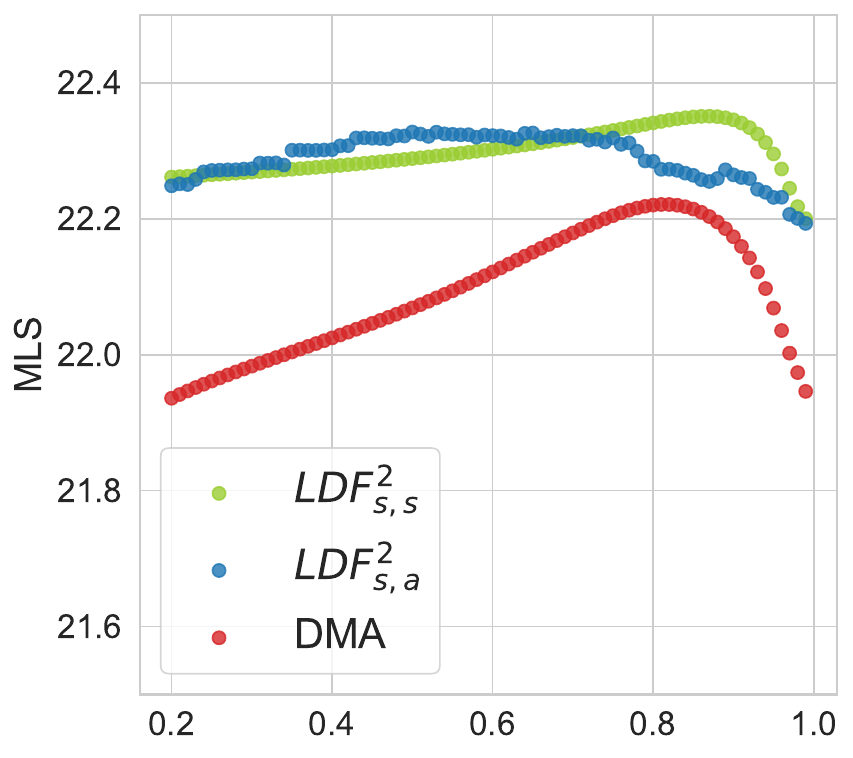}
         \caption{MLS versus values of $\alpha$ for LDF and  for DMA in the x-axis, for the large model pool.}
         \label{fig:fxalphadelta_large_dense}
     \end{subfigure}
\caption{Reconstruction of the top row of Figure \ref{fig:fxalphadelta} with dense grid of parameters $\alpha$ for model averaging.}
        \label{fig:fxalphadelta_dense}
\end{figure}

In Figure \ref{fig:fxalphadelta_dense} we show that once we make the grid of allowable values dense (both between 0.2 and 1 with a space of 0.01) the mean log-scores become smooth functions of the hyperparameters. Naturally a denser grid leads to a slower computation and a such a user can priorities between the speed of calculation and model specification.

\subsection{Results table}
In Table \ref{tab:fx} we show the average and the sum of predictive log-scores for the large and small universes of models for model averaging and selection using LDF methodology, DMA, BMA, BPS, best-N models and random walk benchmark. We see that $LDF_{\mbox{s}, \mbox{s}}^2$ outperforms in model averaging and $LDF_{\mbox{a}, \mbox{s}}^2$ is best in model selection.

\begin{table}
\centering
\resizebox{1.0\textwidth}{!}{%
\input{tables/table_fx.tex}
}
\caption{Predictive log-likelihood for our FX empirical study for a large pool of 2048 models and a small pool of 32 models. We denote the mean log score as $\mbox{MLS}$ and the cumulative log score as $\sum \log(p)$. ``Best'' is the single model which performed best and was was selected a posteriori after we got the results.}
\label{tab:fx}
\end{table}

\subsection{Alternative specifications of BPS}
The Bayesian Predictive synthesis model was run with the following sets of configurations for the FX and corresponding outcomes for the FX study. Each run was performed using 8000 Monte Carlo draws from which 3000 were used for the burnin. We used parallel computing with 34 processes on Intel(R) Xeon(R) Gold 6140 CPU @ 2.30GHz. Each BPS run took around 14 hours. 
\begin{itemize}
    \item $\beta = 0.99$, $\delta = 0.95$, $C_0 = \text{diag}(0.297, \ldots, 0.297)$, $s_0=0.0222$: resulting in MLS = 7.66
    \item $\beta = 0.97$, $\delta = 0.99$, $C_0 = \text{diag}(0.297, \ldots, 0.297)$, $s_0=0.0222$: resulting in MLS = 20.56
    \item $\beta = 0.95$, $\delta = 0.99$, $C_0 = \text{diag}(297, \ldots, 297)$, $s_0=0.0222$: resulting in MLS = 19.02
    \item $\beta = 0.95$, $\delta = 0.99$, $C_0 = \text{diag}(0.297, \ldots, 0.297)$, $s_0=0.0222$: resulting in MLS = 21.60
    \item $\beta = 0.99$, $\delta = 0.99$, $C_0 = \text{diag}(0.297, \ldots, 0.297)$, $s_0=0.0222$: resulting in MLS = 15.94

\end{itemize}

\subsection{GBP/USD case study}
\begin{figure}[!tb]    
\begin{tabular}{cc}
LDF ($\alpha = 0.8$) & BPS\\
        \includegraphics[width=0.47\textwidth]{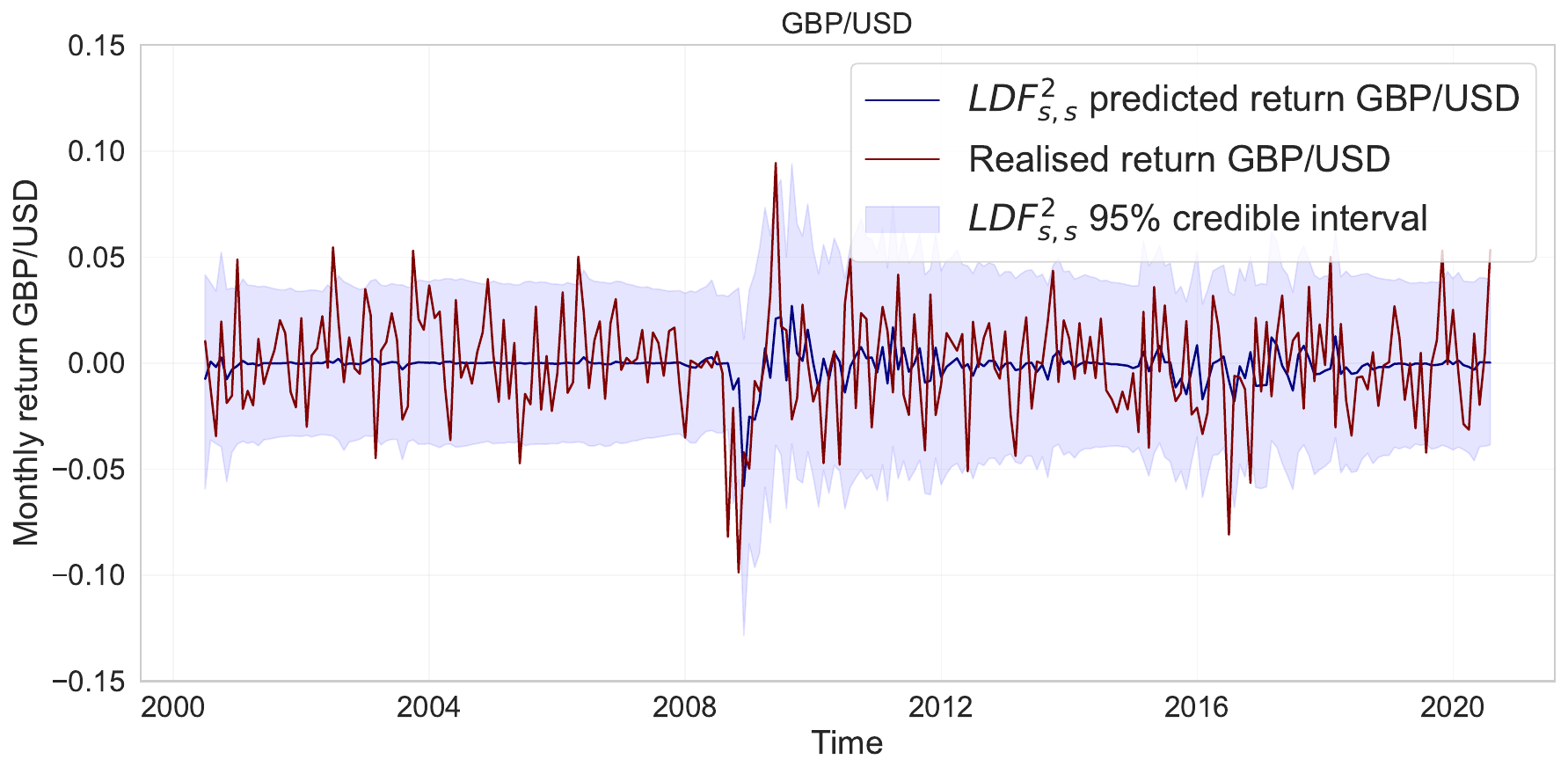}
&
        \includegraphics[width=0.47\textwidth]{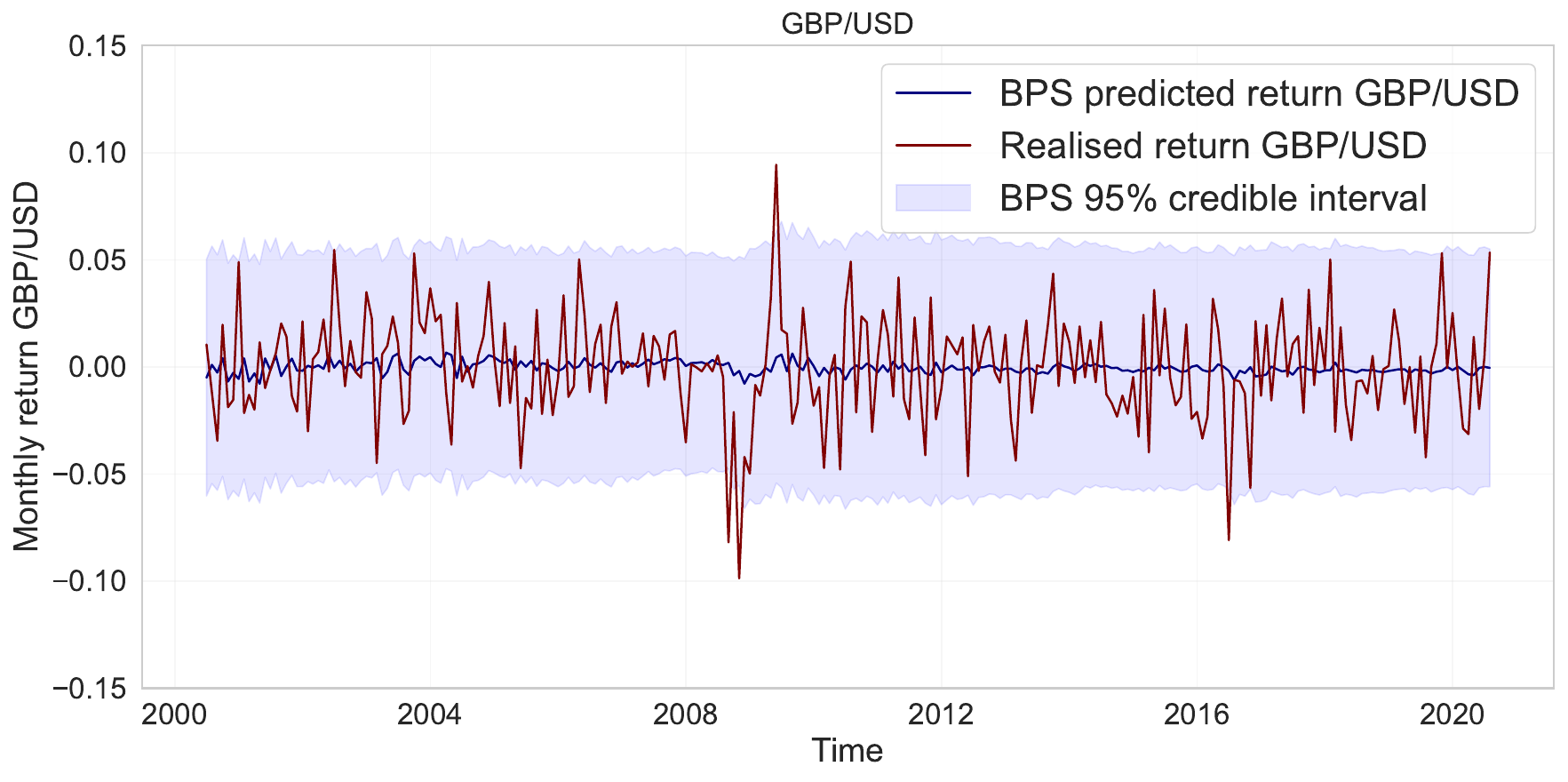}
         \end{tabular}

  \caption{FX: Return forecasts (with 95\% credible intervals) for GBP/USD using LDF and BPS with the small model pool. 
  We can spot that BPS is over-smoothing in this example, whereas LDF model averaging is capable of dynamic adjustments in modelled volatility and returns.
  LDF model averaging with $\alpha=0.8$ modelled returns and 95\% credible intervals versus the realised returns for GBP/USD currency pair.
  BPS modelled returns and 95\% credible intervals versus the realised returns for GBP/USD currency pair.
 }
        \label{fig:fxgbpusd}
\end{figure}

In Figure \ref{fig:fxgbpusd} we show that, for example, for GBP/USD currency rate BPS method showed little directionality as opposed to the $\mbox{LDF}_{\mbox{s}, \mbox{s}}^2$ method. We see that the BPS method ``over-smoothed'' the individual forecasts by giving both the less responsive estimates of mean returns as well as volatility.

\subsection{Data description}
\label{a:data}
The data which was used for the empirical model evaluation was sourced mostly from Refinitiv Eikon, using tickers outlined in Table \ref{tab:data}. There we also show if the data was downloaded from any other sources due to limited history available from our primary source. For EUR we backfilled the history pre 1st January 1999 with German mark exchange rates re-indexed with the conversion rate of $1.95583$. We acknowledge and thank UCL for making the data licence available to support this research. We publish detailed description of the data sources in hope that they can make the results of this research easily replicable.

\afterpage{%
    \clearpage
    \thispagestyle{empty}
    \begin{landscape}

\begin{table}[!htb]
\resizebox{0.85 \textwidth}{!}{%
\begin{tabular}{l|p{3cm}|p{6cm}|p{6cm}|p{6cm}}
Currency    & FX                                  & IR Deposit                                                                                            & IR 10Y Government Bond benchamrk                                                                   & Stock                                                                               \\ \hline \hline
DEM & Bank Of England EoM fixing &                                                                                                       &                                                                                                    &                                                                                     \\
AUD & Eikon                               & From 09/1988 \textit{AUD1MD}, before Australia Overnight Cash rate from archives   of Royal Bank of Australia  & From 06/1990 \textit{AU10YT=RR}, Before Fed St Luis                                                         & S\&P 200 \textit{.AXJO} From 05/1992 before All ordinaries Index                             \\
CAD & Eikon                               & \textit{CAD1MD}& From 06/1986 \textit{CA10YT=RR}, before FED St Luis                                                         & S\&P Comp Index \textit{.GSPTSE}                                                        \\
EUR & Eikon                               & From 01/1999 onwards \textit{EUR1MD}, From 02/1994 to 12/1998 DE1MT Eikon, Before Bundesbank Monthly average & From 01/1991 \textit{EU10YT=RR}, From 08/1992 to 12/1998 DE10YT, before Germany   10Y Benchmark FED St Luis & DAX \textit{.GDAXI} From 12/1987                                                             \\
JPY & Eikon                               & \textit{JPY1MD}                                                                                                & \textit{JP10YT=RR} & NIKKEI 225 \textit{.N225E} \\
NZD & Eikon                               & From 09/1988 \textit{NZD1MD}, before New Zealand central bank New Zealand stats   wholesale 30d bills          & From -1/1996 \textit{NZ10YT=RR}, before Fed St Luis                                                         & NZX 50 from 12/2000, Before NZX All rebased to NZX 50 \textit{.NZCI}, first date   06/1986 \\
NOK & Eikon                               & From 09/1988 \textit{NOK1MD}, Bank of Norway overnight monthly rate                                            & From 03/1994 \textit{NO10YT=RR}, Before Fed St Luis                                                         & \textit{OBX} From 09/1999, before \textit{OSEAX} rebased                                              \\
SEK & Eikon                               & From 09/1988 \textit{SEK1MD}, Fed Bank St Luis average monthly rates                                           & From 05/1991 \textit{SE10YT=RR}, Before Fed St Luis                                                         & OMXS 30 From 09/1986                                                                \\
CHF & Eikon                               & \textit{CHF1MD}  & From 01/1992 \textit{CH10YT=RR}, Before Fed St Luis                                                         & SMI \textit{.SSMI} From 01/1988                                                            \\
GBP & Eikon                               & \textit{GBP1MD}                                                                                                & From 01/1990 \textit{GB10YT=RR}, Before Fed St Luis                                                         & FTSE100 \textit{.FTSE}                                                                     \\
USD & Eikon                               & \textit{USD1MD}                                                                                                & From 08/1987 \textit{US10YT=RR}, Before Fed St Luis                                                         & S\&P 500 \textit{.SPX}                                                                    
\end{tabular}%
}
\caption{Sources of data used for FX study.}
\label{tab:data}
\end{table}

   \end{landscape}
    \clearpage
}

%% file: tables/table_fx.tex
\begin{tabular}{c|c|c||c|c|c||c|c|c||c|c|c}
\multicolumn{6}{c||}{Large model pool} & \multicolumn{6}{c}{Small model pool}\\ \hline \hline
\multicolumn{3}{c||}{Model Averaging}   & \multicolumn{3}{c||}{Model Selection} & \multicolumn{3}{c||}{Model Averaging}   & \multicolumn{3}{c}{Model Selection} \\ \hline 
\textbf{Model} & \textbf{MLS} & \textbf{$\sum \log(p)$}       & \textbf{Model}               & \textbf{MLS}        & \textbf{$\sum \log(p)$}  & \textbf{Model} & \textbf{MLS} & \textbf{$\sum \log(p)$}       & \textbf{Model}               & \textbf{MLS}        & \textbf{$\sum log(p)$}   \\ \hline \hline
\multicolumn{3}{c||}{$\mbox{LDF}_{\mbox{s}, \mbox{a}}^2$}   & \multicolumn{3}{c||}{$\mbox{LDF}_{\mbox{a}, \mbox{a}}^2$} & \multicolumn{3}{c||}{$\mbox{LDF}_{\mbox{s}, \mbox{a}}^2$}   & \multicolumn{3}{c}{$\mbox{LDF}_{\mbox{a}, \mbox{a}}^2$}\\ \hline
$\alpha=1.00$   & 22.17      & 5367.23             & $\alpha=1.00$                  & 21.75               & 5263.83        & $\alpha=1.00$ & 22.04 & 5334.29 & $\alpha=1.00$ & 22.04 & 5333.80 \\
$\alpha=0.95$   & 22.22      & 5376.89             & $\alpha=0.95$                  & 21.78               & 5270.54        & $\alpha=0.95$ & 22.09 & 5346.56 & $\alpha=0.95$ & 22.06 & 5334.12 \\
$\alpha=0.90$   & 22.24      & 5381.82             & $\alpha=0.90$                  & 21.80               & 5275.55        & $\alpha=0.90$ & 22.11 & 5351.78 & $\alpha=0.90$ & 22.06 & 5339.27 \\
$\alpha=0.80$   & 22.27      & 5389.48             & $\alpha=0.80$                  & 21.75               & 5263.94        & $\alpha=0.80$ & 22.14 & 5358.06 & $\alpha=0.80$ & 22.05 & 5338.22  \\
$\alpha=0.70$   & 22.32      & 5400.33             & $\alpha=0.70$                  & 21.73               & 5259.77        & $\alpha=0.70$ & 22.13 & 5355.35 & $\alpha=0.70$ & 22.00 & 5335.29     \\ \hline \hline
\multicolumn{3}{c||}{$\mbox{LDF}_{\mbox{s}, \mbox{s}}^2$}   & \multicolumn{3}{c||}{$\mbox{LDF}_{\mbox{a}, \mbox{s}}^2$} & \multicolumn{3}{c||}{$\mbox{LDF}_{\mbox{s}, \mbox{s}}^2$}   & \multicolumn{3}{c}{$\mbox{LDF}_{\mbox{a}, \mbox{s}}^2$}\\ \hline
$\alpha=1.00$   & 22.17      & 5365.09             & $\alpha=1.00$                  & 21.77               & 5268.73        & $\alpha=1.00$ & 22.04 & 5334.40 & $\alpha=1.00$ & 22.04 & 5333.80 \\
$\alpha=0.95$   & 22.30      & 5396.55             & $\alpha=0.95$                  & 21.84               & 5319.99        & $\alpha=0.95$ & 22.13 & 5355.42 & $\alpha=0.95$ & 22.11 & 5351.02 \\
$\alpha=0.90$   & 22.36      & 5411.00             & $\alpha=0.90$                  & 21.98               & 5330.19        & $\alpha=0.90$ & 22.16 & 5362.06 & $\alpha=0.90$ & 22.14 & 5357.23 \\
$\alpha=0.80$   & 22.37      & 5413.55             & $\alpha=0.80$                  & 22.03               & 5332.49        & $\alpha=0.80$ & 22.15 & 5359.58 & $\alpha=0.80$ & 22.13 & 5355.84  \\
$\alpha=0.70$   & 22.35      & 5409.71             & $\alpha=0.70$                  & 22.03               & 5331.62        & $\alpha=0.70$ & 22.13 & 5354.71 & $\alpha=0.70$ & 22.12 & 5353.05     \\ \hline \hline
\multicolumn{3}{c||}{\textbf{Dynamic Model Averaging}}   & \multicolumn{3}{c||}{$\mbox{LDF}_{\mbox{a}}^1$} & \multicolumn{3}{c||}{\textbf{Dynamic Model Averaging}}   & \multicolumn{3}{c}{$\mbox{LDF}_{\mbox{a}}^1$}\\ \hline
$\alpha=0.95$   & 22.07      & 5340.61             & $\alpha=0.95$                 & 21.85               & 5287.76         & $\alpha=0.95$ & 22.04 & 5334.64 & $\alpha=0.95$ & 21.93 & 5307.47 \\
$\alpha=0.90$   & 22.17      & 5366.03             & $\alpha=0.90$                 & 21.81               & 5277.69         & $\alpha=0.90$ & 22.11 & 5350.37 & $\alpha=0.90$ & 21.96 & 5314.51 \\
$\alpha=0.80$   & 22.22      & 5377.29             & $\alpha=0.80$                 & 21.78               & 5269.78         & $\alpha=0.80$ & 22.08 & 5344.33 & $\alpha=0.80$ & 22.01 & 5326.75  \\
$\alpha=0.70$   & 22.18      & 5367.38             & $\alpha=0.70$                 & 21.77               & 5268.73         & $\alpha=0.70$ & 22.01 & 5326.51 & $\alpha=0.70$ & 22.04 & 5333.80    \\ \hline \hline
Average         & 20.82      & 5037.57             & RW                            & 21.77               & 5267.81         & BPS           & 21.60 & 5227.56 & Average      & 21.71 & 5253.55   \\
BMA             & 21.91      & 5302.83             & Best                          & 21.78               & 5271.02         & Best-4 avg.   & 22.10 & 5349.11 & Best         & 21.77 & 5267.81     \\
\end{tabular}%

%% file: parts/appendices/appendix_usinflation.tex
\section{US inflation study - supplementary material}
\label{appendix:E}    
\begin{figure}[!tb]
     \centering
 \includegraphics[width=0.8\linewidth]{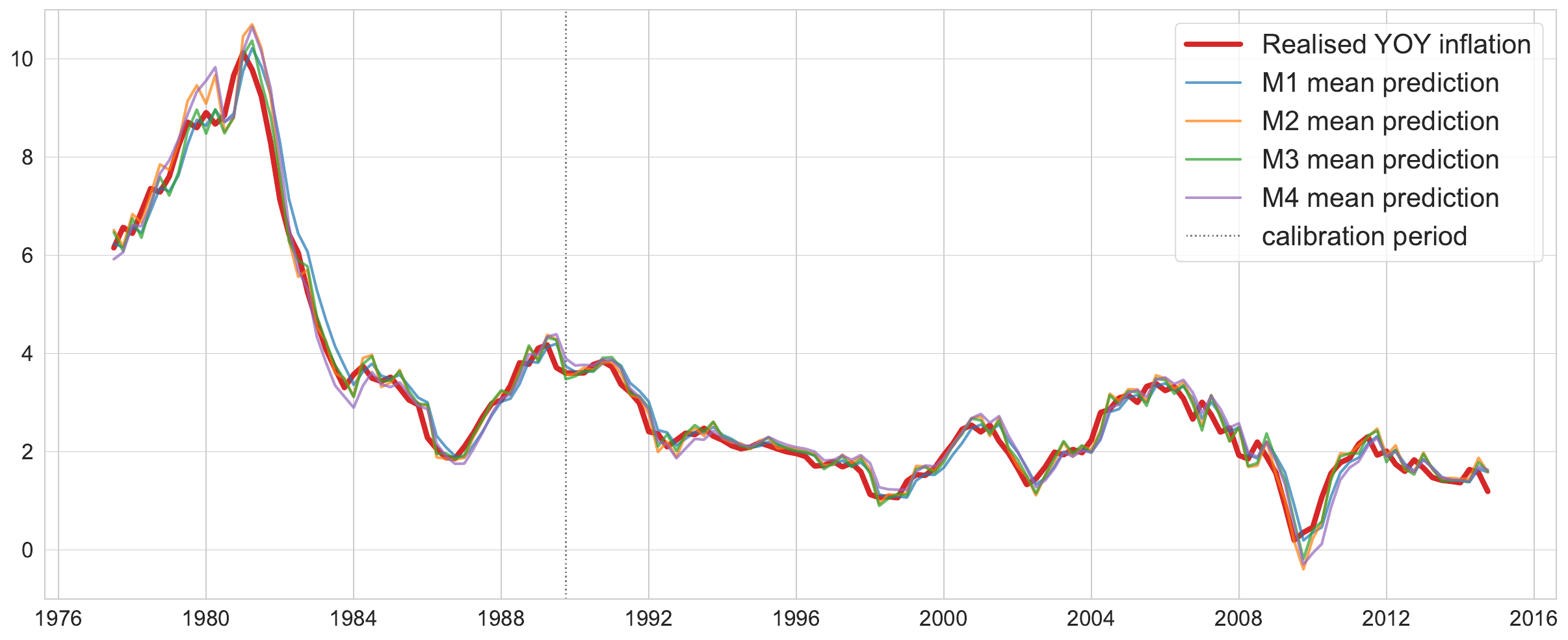}  
 \label{fig:sub-first}
\caption{US inflation: Forecast means for each model and realised year-on-year 
}
\label{fig:usinf}
\end{figure}

In Figure \ref{fig:usinf} we show the forecast means of each of four models considered vs realised year-on-year inflation.

In Table \ref{tab:usinf} we show the full set of results for this study.

\begin{table}
\centering
\resizebox{0.7\textwidth}{!}{%
\input{tables/table_usinf}
}
\caption{Predictive log-likelihood for US inflation example from \citet{McAlinn_2019}. We denote the mean log score as $\mbox{MLS}$, the cumulative log score as $\sum \log(p)$. We note that BPS model provides the best synthesis in this example}
\label{tab:usinf}
\end{table}

%% file: tables/table_usinf.tex
\begin{tabular}{c|c|c||c|c|c}
\textbf{Model}    & $MLS$   & $\sum \log(p)$ & \textbf{Model}  & $MLS$& $\sum \log(p)$  \\
\hline
\hline
BMA               & -0.03           & -2.97	        & BPS               & 0.06           & 6.10 \\
\hline 
\multicolumn{3}{c||}{\textbf{Best N-average, $rw=20$}} & \multicolumn{3}{c}{\textbf{Single models}}  \\
\hline
N=1               & -0.06           & -5.75        & $M_1$            & -0.08           & -7.77      \\
N=2               & -0.01           & -0.87	       & $M_2$            & -0.02           & -2.48      \\
N=3               & -0.01           & -1.40        & $M_3$            & -0.03           & -2.97      \\
N=4               & -0.03           & -3.03        & $M_4$            & -0.17           & -16.65      \\
\hline
\multicolumn{3}{c||}{$LDF_{\mbox{s}, \mbox{s}}^2$} & \multicolumn{3}{c}{\textbf{Dynamic Model Averaging}}                               \\
\hline
$\alpha =1.00$   & -0.03           & -2.76                & $\alpha = 1.00$   & -0.03           & -2.97	    \\
$\alpha =0.95$   & -0.03           & -2.29                & $\alpha = 0.95$   & -0.02           & -2.40	    \\
$\alpha =0.90$   & -0.04           & -2.47                & $\alpha = 0.90$   & -0.03           & -2.73	    \\
$\alpha =0.80$   & -0.04           & -2.55                & $\alpha = 0.80$   & -0.03           & -3.07	    \\
$\alpha =0.70$   & -0.04           & -2.57                & $\alpha = 0.70$   & -0.03           & -3.11                            
\end{tabular}%